\documentclass[twocolumn,english,aip,amsmath,amssymb,nolinenumbers,reprint]{revtex4-1}
\usepackage[T1]{fontenc}
\usepackage[utf8]{inputenc}
\usepackage{color}
\usepackage{babel}
\usepackage{amsmath}
\usepackage{amssymb}
\usepackage{graphicx}
\usepackage[unicode=true,pdfusetitle,
 bookmarks=true,bookmarksnumbered=false,bookmarksopen=false,
 breaklinks=false,pdfborder={0 0 1},backref=false,colorlinks=true]
 {hyperref}
\hypersetup{pdfborderstyle=,pdfborderstyle={},citecolor=blue}

\makeatletter

\usepackage{color}
\usepackage{babel}

 \graphicspath{{./}{./FIGURES/}} 
 
 \newcommand{\biftool}[1]{\texttt{dde-biftool}}

\makeatother

\begin{document}
\title{Topological Localized states the time delayed Adler model: Bifurcation
analysis and interaction law}
\author{L. Munsberg }
\affiliation{Institute for Theoretical Physics, University of Münster, Wilhelm-Klemm-Str. 9,
D-48149 Münster, Germany}
\affiliation{Center for Nonlinear Science (CeNoS), University of Münster, Corrensstrasse
2, D-48149 Münster, Germany}
\author{J. Javaloyes}
\affiliation{Departament de Física \& IAC-3, Universitat de les Illes Balears,
C/ Valldemossa km 7.5, 07122 Mallorca, Spain}
\author{S. V. Gurevich}
\email{gurevics@uni-muenster.de}

\affiliation{Institute for Theoretical Physics, University of Münster, Wilhelm-Klemm-Str. 9,
D-48149 Münster, Germany}
\affiliation{Center for Nonlinear Science (CeNoS), University of Münster, Corrensstrasse
2, D-48149 Münster, Germany}
\affiliation{Departament de Física \& IAC-3, Universitat de les Illes Balears,
C/ Valldemossa km 7.5, 07122 Mallorca, Spain}
\begin{abstract}
The time-delayed Adler equation is arguably the simplest model for
an injected semiconductor laser with coherent injection and optical
feedback. It is able to reproduce the existence of topological localized
structures (LSs) and their rich interactions. In this paper we perform
the first extended bifurcation analysis of this model and we explore
the mechanisms by which LSs emerge. We also derive the effective equations
governing the motion of distant LSs and we stress how the lack of
parity in time-delayed systems leads to exotic, non-reciprocal, interactions
between topological localized states.
\end{abstract}
\maketitle

\section{Introduction}

Localized structures (LSs) appear in driven dissipative nonlinear
systems and they can be observed in a variety of complex systems \citep{WKR-PRL-84,MFS-PRA-87,NAD-PSS-92,UMS-NAT-96,AP-PLA-01}.
They are \emph{attractors} of the dynamics, i.e. stable solutions
towards which the system converges from a wide set of initial conditions
\citep{NP-SelfOrg-77}. In Optics, LSs are usually envisioned as light
pulses in time or localized beams in space, see \citep{L-CSF-94,MT-JOSAB-04,AFO-AAM-09}
for reviews, and one distinguishes between systems in which the LSs
are locked to an external injection beam from the ones that possess
a phase invariance. The former situation leads to the so-called cavity
solitons \citep{FS-PRL-96,BLP-PRL-97} observed either in the transverse
plane of broad area amplifiers \citep{BTB-NAT-02} or in the temporal
output of fibers \citep{LCK-NAP-10,HBJ-NAP-14}. In phase invariant
situations, spatial diffractive autosolitons were predicted and observed
either in cavities composed of a gain medium coupled to a saturable
absorber \citep{RK-OS-88,RK-JOSAB-90,GBG-PRL-10} or to an external
diffraction grating \citep{TAF-PRL-08}. Temporal localization was
also achieved in passively mode-locked lasers operated in the long
cavity regime \citep{MJB-PRL-14} and it was demonstrated that a similar
setup could also lead to full thee-dimensional spatio-temporal localization
\citep{J-PRL-16}.

Spatial and temporal LSs of light have often been analyzed using similar
theoretical frameworks. However, while space is isotropic, temporal
dynamics usually exhibits a symmetry breaking due to the causal response
of the active medium. The dynamics of temporal LSs may break the action-reaction
principle, an effect that was discussed for instance in the framework
of mode-locking \citep{JCM-PRL-16,CJM-PRA-16}. Building upon the
strong analogies between spatially extended and time-delayed systems
\citep{AGL-PRA-92,GP-PRL-96,K-CMMP-98} (TDSs), the latter have been
proposed for \emph{generating} temporal LSs, see \citep{YG-JPA-17}
for a recent review. In TDSs, propagation and nonlinearity occur in
well separated stages, which is at variance with distributed systems,
such as, e.g., the nonlinear Schrödinger equation governing light
propagation in fibers. In most cases, LSs appear in TDSs as mutually
independent light peaks \citep{MJB-PRL-14,RAF-SR-16}. However, topological
LSs were also predicted and observed in semiconductor lasers; They
exist either as $2\pi$ kinks in the polarization orientation \citep{MJB-NAP-15}
or in the phase of the optical field \citep{GJT-NC-15}. In this latter
case, it was shown that a simple time-delayed model for the phase
of the lasing field is able to reproduce the results obtained in \citep{GJT-NC-15,GJB-CHA-17}.
Despite its simplicity, this model contains the effects of optical
injection, frequency locking and time-delayed feedback and can be
termed \emph{the time-delayed Adler equation}.

In this manuscript, we perform an extended analysis of the delayed
Adler equation using the path continuation package \texttt{dde-biftool}
\citep{DDEBT} and asymptotic analysis. In particular, we provide
the effective equations of motion for distant, weakly interacting,
topological LSs and we give the conditions under which repulsive and
attractive forces can give rise to stable molecules. Finally, we show
that the interaction between LSs are not reciprocal, a feature typical
of system with broken parity symmetry.

\section{Model}

Topological LSs can be obtained by combining two elements. The first
is a semiconductor laser with coherent optical injection operated
in the so-called ``excitable'' regime \citep{GDP-PRE-17}. In this
regime, the phase of the semiconductor laser is stably locked to the
external forcing. Upon small perturbations, the phase, that evolve
on a circle, relaxes (e.g., clockwise) exponentially to its equilibrium
state. However, when responding to a sufficiently large external perturbation,
the phase performs a $2\pi$ anti-clockwise rotation, after which
the system locks again to the external forcing. This mechanism leads
to a simple scenario of excitability. To this injected laser system,
we add a delayed feedback loop. Here, the delayed feedback plays the
role of the extended (spatial) degree of freedom \citep{GP-PRL-96}
in which multiple independently addressable topological LSs can be
stored and regenerated indefinitely \citep{GJT-NC-15}. Each $2\pi$
phase structures, embedded in a homogeneously locked background field,
propagate into the external feedback loop, similarly to the sine-Gordon
solitons\citep{CDT-PRE-98}. When these kinks come back into the semiconductor
laser they act as triggers for new excitable dynamics and, hence they
get regenerated. By assuming small injection and optical feedback,
as well as a small detuning between the injection field and the laser
natural frequency, a multiple time scale analysis yields the time-delayed
Adler equation for the phase evolution $\theta\left(t\right)$
\begin{eqnarray}
\dot{\theta} & = & \Delta-\sin\theta+\chi\sin\left[\theta\left(t-\tau\right)-\theta-\psi\right]\,,\label{eq:adler}
\end{eqnarray}
where the dot denotes the time derivative with respect to a slow time
$t$, $\Delta$ the ratio and of the detuning between the injection
and the laser normalized to the injection field amplitude, $\chi>0$
the ratio of the feedback rate and of the amplitude of the injection
field, while the parameter $\psi$ is directly related to the feedback
phase, see Sup. Mat. in \citep{GJT-NC-15} for more details. In the
absence of feedback $\left(\chi=0\right)$, Eq.~\eqref{eq:adler}
becomes an Adler equation that describes e.g., a time evolution of
the phase difference of two weakly coupled oscillators with a small
detuning in their frequencies. Here, one notices the presence of saddle
node on a circle bifurcation that arises at $\Delta_{\pm}=\pm1$;
It is in the vicinity of $\Delta_{\pm}$ that the Adler equation exhibits
excitability. Temporal LSs are found in the long time delay limit
\citep{YRS-PRL-19} in which the excitable orbit duration is shorter
than the time delay. From the strict point of view of TDSs, temporal
LSs are peculiar periodic solutions that consists in a localized waveform,
that is independent of the precise value of $\tau$, if $\tau$ is
large enough, embedded into an arbitrary large chunk of the homogeneous
solution. Recently, a complete classification of the spectrum of TDSs
into interface and pseudo-continuous spectrum was achieved \citep{YRS-PRL-19}
and the link between LSs periodic orbits with homoclinic solutions
was clarified.

\section{Results}

We start the analysis by searching the steady states of Eq.~\eqref{eq:adler}
that are defined by
\begin{align*}
0= & \Delta-\sin\theta-\chi\sin\psi,
\end{align*}
which is solved by
\begin{align*}
\theta_{s}= & \arcsin\left(\Delta-\chi\sin\psi\right)+2\pi n, & n\in\mathbb{Z},\\
\theta_{u}= & \pi-\arcsin\left(\Delta-\chi\sin\psi\right)+2\pi n. & n\in\mathbb{Z}.
\end{align*}
making that $\theta_{s}$ (resp. $\theta_{u}$) belongs to the right
(resp. left) half of the unit circle, i.e. $\theta_{s}\in\left[-\frac{\pi}{2},\,\frac{\pi}{2}\right]$
and $\theta_{u}\in\left[\frac{\pi}{2},\,\frac{3\pi}{2}\right]$. The
linear stability analysis information for the surroundings of the
steady states is obtained setting $\theta_{s,u}+\varepsilon\delta$
with $\varepsilon\ll1$, which yields 
\begin{align}
\dot{\delta}= & A\delta(t)+B\delta(t-\tau),\\
A= & -\cos\theta_{s,u}-\chi\cos\psi,\label{eq:A}\\
B= & \chi\cos\psi\,,\label{eq:B}
\end{align}
where $A$ and $B$ are the instantaneous and delayed Jacobian matrices
of the linearized problem evaluated at the steady state. The characteristic
equation is found setting $\delta\left(t\right)=\delta_{0}\exp\left(\lambda t\right)$,
leading to the transcendental equation
\begin{align}
\lambda= & A+Be^{-\lambda\tau}.\label{eq:ststeigenvalues}
\end{align}

Note that without delayed feedback ($\chi=0$) the eigenvalues $\lambda_{s,u}$
are given by: 
\begin{align}
\lambda_{s,u} = & -\cos\theta_{s,u},
\end{align}
leading to $\lambda_{s}<0$ and $\lambda_{u}>0$. Therefore without
delay the solution $\theta_{s}$ is always stable and $\theta_{u}$
is always unstable. Furthermore, for $\Delta_{\pm}=\pm1$ both steady
states coincide at: 
\begin{align*}
\theta_{\pm}= & \pm\frac{\pi}{2}+2\pi n, & n\in\mathbb{Z}
\end{align*}
and disappear in a saddle-node bifurcation for $|\Delta|>1$.

Considering the influence of the delay, the steady states only exist
for: 
\begin{align*}
\Delta-\chi\sin\psi\in & [-1,\,1]
\end{align*}
leading to the boundaries of existence for the steady states as 
\begin{align}
\Delta_{sn}^{+}= & 1+\chi\sin\psi,\label{eq:saddlenode1}\\
\Delta_{sn}^{-}= & -1+\chi\sin\psi,\label{eq:saddlenode2}
\end{align}
which effectively shifts the saddle-node bifurcation to higher values
of $\Delta$ for $\psi\in[0,\,\pi]$ and to lower values for $\psi\in[\pi,\,2\pi]$.
The amplitude of the shift is given by $\chi$.

\begin{figure}[ht!]
\includegraphics[width=0.49\columnwidth]{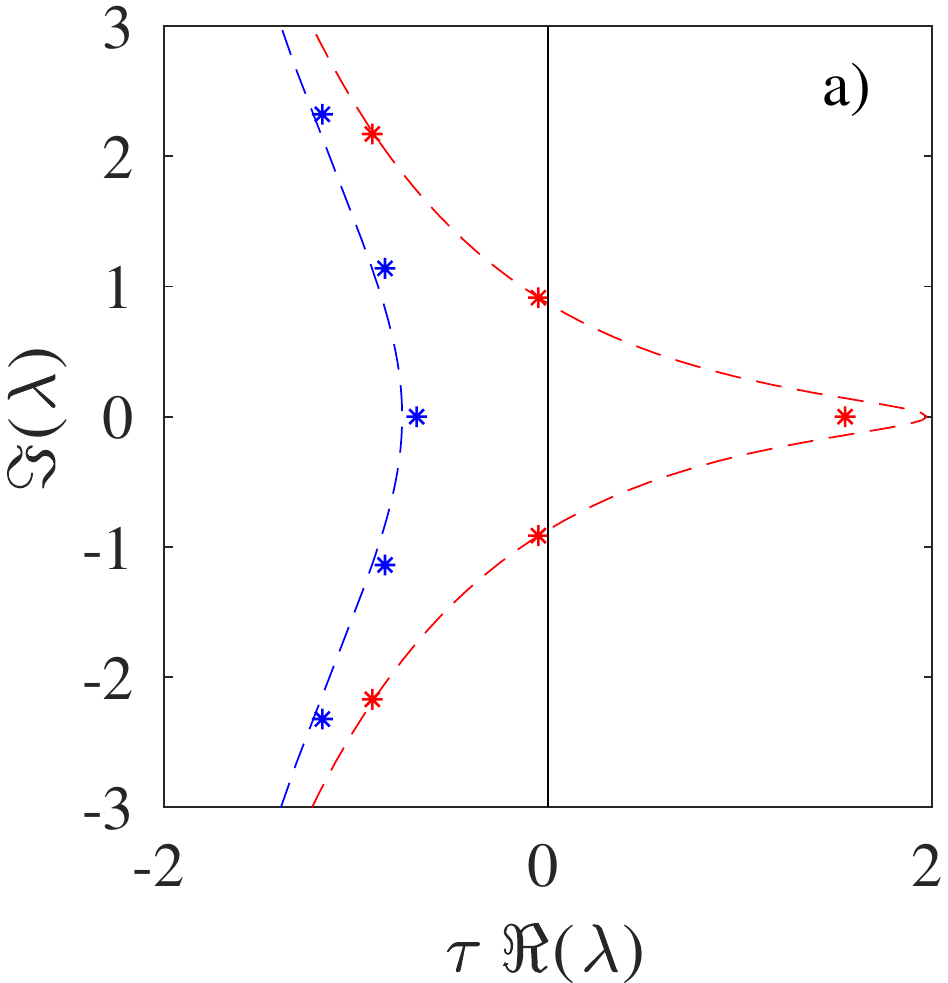} \includegraphics[width=0.49\columnwidth]{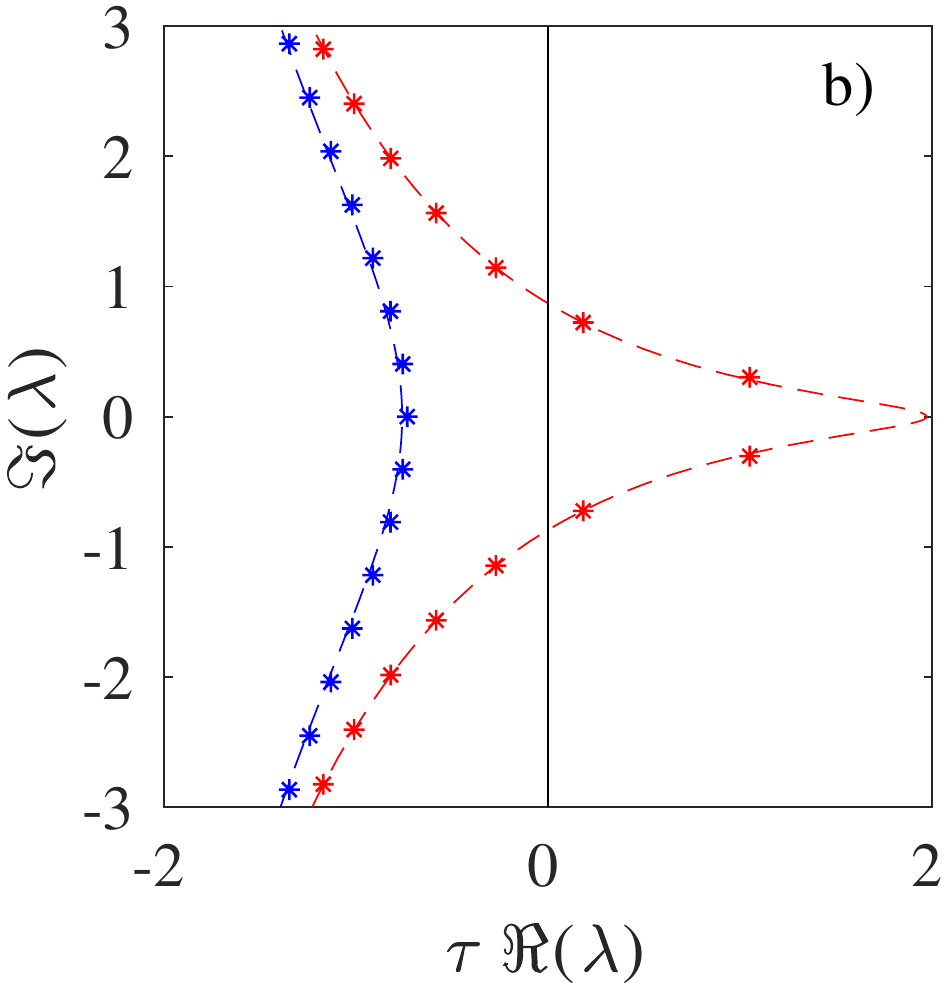}
\caption{Eigenvalues $\lambda_{s}$ (blue) and $\lambda_{u}$ (red) of Eq.~\eqref{eq:ststeigenvalues}
corresponding to the steady states $\theta_{s,u}$ for (a) $\tau=5$
and (b) $\tau=15$. The dashed lines show the quasi-continuous spectrum
\eqref{eq:pseudocont} while the stars represent the exact solutions
given by the Lambert W functions \eqref{eq:lambert}. Other parameters
are $\left(\Delta,\chi,\psi\right)=\left(0.5,1,0.5\right)$.}
\label{fig:eigenvalues} 
\end{figure}

The characteristic equation \eqref{eq:ststeigenvalues} can not be
solved analytically for $\chi\neq0$ as it is transcendental. However
solutions for $\lambda$ can be found using the Lambert $W$ functions\citep{CGHJK-ACM-96}
\begin{align}
\lambda_{n}= & A+\frac{1}{\tau}W_{n}\left(B\tau e^{-A\tau}\right).\label{eq:lambert}
\end{align}
The resulting number of eigenvalues is infinite because of the infinite
number of branches of the Lambert $W$ functions denoted $W_{n}$
with $n\in\mathbb{Z}$. In the limit of long delays the infinite number
of eigenvalues accumulate over a quasi-continuous spectrum \citep{Y-DCDS-15}.
By expanding the eigenvalues in real and imaginary part $\lambda=\frac{\alpha}{\tau}+i\beta$
in Eq.~\eqref{eq:ststeigenvalues} one obtains
\begin{align}
\alpha= & \frac{1}{2}\ln\left(\frac{B^{2}}{A^{2}+\beta^{2}}\right).\label{eq:pseudocont}
\end{align}
In figure \ref{fig:eigenvalues} the quasi-continuous spectrum as
well as the leading eigenvalues obtained from Eq.~\eqref{eq:lambert}
are plotted. One can observe that even for small values of the delay
time $\tau=5$ (see the panel (a)), the quasi-continuous spectrum
is a good approximation for the exact eigenvalues obtained from Eq.~\eqref{eq:lambert}.
For larger values of $\tau$ the distance between the discrete eigenvalues
is smaller leading to an increasing number of relevant eigenvalues.

For large delays one can estimate the instability threshold for the
steady states by calculating the set of parameters leading to a crossing
of the quasi-continuous spectrum with the imaginary axis. Depending
on whether the leading discrete eigenvalues are real or a set of complex
conjugates, the occurring bifurcation is a saddle-node (SN) or Andronov-Hopf
(AH) bifurcation. In our case, the first AH point is given by $\alpha=0$
that we approximate in the long delay limit as $\beta\sim1/\tau=0$.
Because the maximum real part of Eq.~\eqref{eq:pseudocont} is obtained
for a vanishing imaginary part, this leads to the relation $A^{2}=B^{2}$.
The first two borders obtained by setting $A=-B$ correspond to the
SN bifurcation identified previously in Eqs.~(\eqref{eq:A},\eqref{eq:B})
while the branch $A=B$ yields the following borders of stability
at which the steady state gets AH unstable; for the solutions $\theta_{s}$
and $\theta_{u}$, these borders have an explicit expression that
reads
\begin{eqnarray}
\Delta_{s}^{\pm} & = & \pm\sqrt{1-4\chi^{2}\cos^{2}\psi}+\chi\sin\psi\,,\cos\psi<0\label{eq:border3}\\
\Delta_{u}^{\pm} & = & \pm\sqrt{1-4\chi^{2}\cos^{2}\psi}+\chi\sin\psi\,,\cos\psi>0\label{eq:border5}
\end{eqnarray}

\begin{figure}[ht!]
\includegraphics[width=0.32\columnwidth]{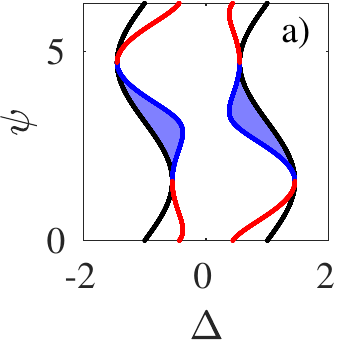}\includegraphics[width=0.32\columnwidth]{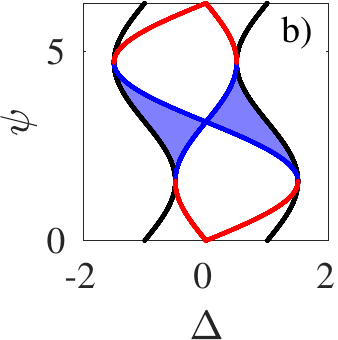}\includegraphics[width=0.32\columnwidth]{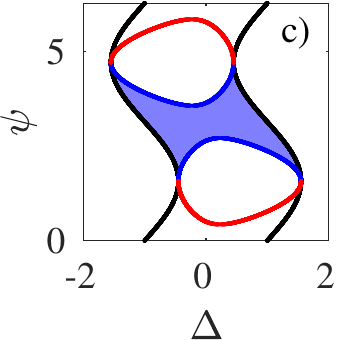}
\caption{Dependence of AH unstable regime, shaded in blue, on feedback strength
$\chi$ in the long delay limit \eqref{eq:pseudocont}. Blue lines
show the AH instabilities of $\theta_{s}$, while red lines show the
AH instabilities of $\theta_{u}$ and black lines correspond to the
SN bifurcation of both steady states. Feedback strength changes from
(a) $\chi=0.45$, (b) $\chi=0.5$ to (c) $\chi=0.55$.}
\label{fig:hopfregionlongdelay} 
\end{figure}

The area in which the steady state $\theta_{s}$ is AH unstable is
especially important because this results in an oscillating background
for LSs. The change of size and shape of the AH unstable area as a
function of the feedback strength $\chi$ was investigated in Fig.
\ref{fig:hopfregionlongdelay}; the four borders resulting from Eqs.~\eqref{eq:border3},\eqref{eq:border5}
are plotted for different values of $\chi$ and the region of instability
of $\theta_{s}$ is colored in blue. For low feedback strengths, there
are two areas in which the steady state $\theta_{s}$ is AH unstable
and they are close to the SN bifurcation between $\psi=\pi/2$ and
$\psi=3\pi/2$, see Fig.~\ref{fig:hopfregionlongdelay}(a). With
increasing feedback strength, these areas stretch further away from
the SN border, until they meet at a critical point defined by $\left(\Delta_{c},\chi_{c},\psi_{c}\right)=\left(0,1/2,\pi\right)$
as shown in Fig.~\ref{fig:hopfregionlongdelay}(b). For even bigger
feedback strengths of $\chi>\chi_{c}$ the two areas merge into one
area of instability (cf. Fig.~\ref{fig:hopfregionlongdelay}(c)).

\begin{figure}[ht!]
\includegraphics[width=0.49\columnwidth]{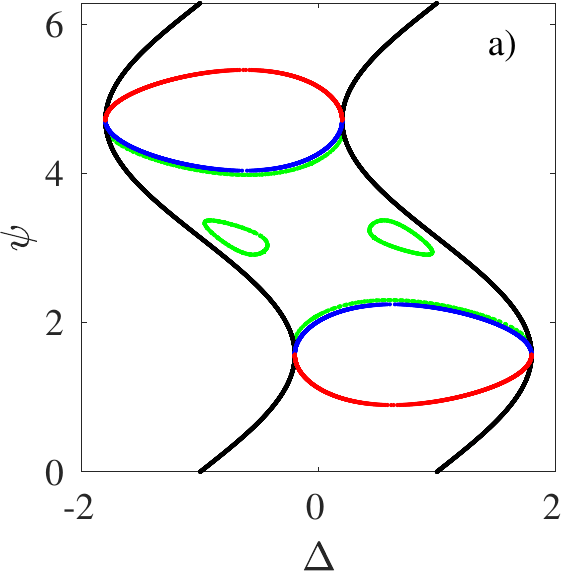}\includegraphics[width=0.49\columnwidth]{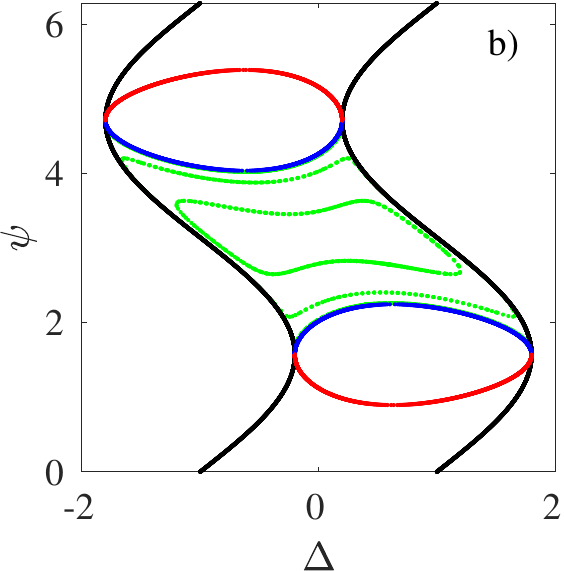}
\caption{Borders of the AH instability obtained from the path-continuation
(green) for (a) $\tau=10$ and (b) $\tau=20$, respectively. Steady
state instabilities in the long delay limit are shown in black (SN)
blue (AH of $\theta_{s}$) and red (AH of $\theta_{u}$). }
\label{fig:hopfcontinuation} 
\end{figure}

Using the \texttt{dde-biftool\citep{DDEBT}} framework it is possible
to follow the steady state $\theta_{s}$ while varying the parameters
$\Delta$ and $\psi$. If there an AH bifurcation occurs, it is possible
to follow the AH bifurcation point in the $\left(\Delta,\psi\right)$
plane. Since several pairs of eigenvalues can cross the imaginary
axis, this method allows obtaining several curves corresponding to
the crossing of the imaginary axis. In Fig. \ref{fig:hopfcontinuation}
the results of the continuation are shown for two different time delays
$\tau$. For each $\tau$ one can observe an AH border close to but
not exactly at the border obtained from the long delay limit approximation,
corresponding to the crossing of the first pair of complex conjugate
eigenvalues with the imaginary axis. For small delay times as shown
in Fig.\ref{fig:hopfcontinuation}(a) for $\tau=10$, the regions
where a second pair of complex conjugate eigenvalues have a positive
real part form two separate small ovals centered around $\psi=\pi$.
With increasing delay time the distance between the discrete eigenvalues
becomes smaller, while they still follow the quasi-continuous spectrum
which is independent on the delay time. This leads to a larger region
of instability for each pair of complex conjugate eigenvalues crossing
the imaginary axis. In the case of Fig.~\ref{fig:hopfcontinuation}(b),
the larger delay $\tau=20$ leads to a merging of the two small regions
of secondary instability into one large region. Notice that a good
approximation of these additional AH lines, and more generally of
the complex eigenvalue spectrum, can be obtained setting $\beta_{n}\sim2\pi\left(n+\frac{1}{2}\right)/\tau$
in Eq.~\eqref{eq:pseudocont} to find $\alpha_{n}$. Setting $\alpha_{n}$
allows finding the green curves depicted in Fig.~\ref{fig:hopfcontinuation}.
We note that the first AH bifurcation appears with frequency $\beta_{1}\sim\pi/\tau$
which corresponds to a period two regime, characteristic of time delayed
systems. 

\begin{figure}[ht!]
\includegraphics[width=0.49\columnwidth]{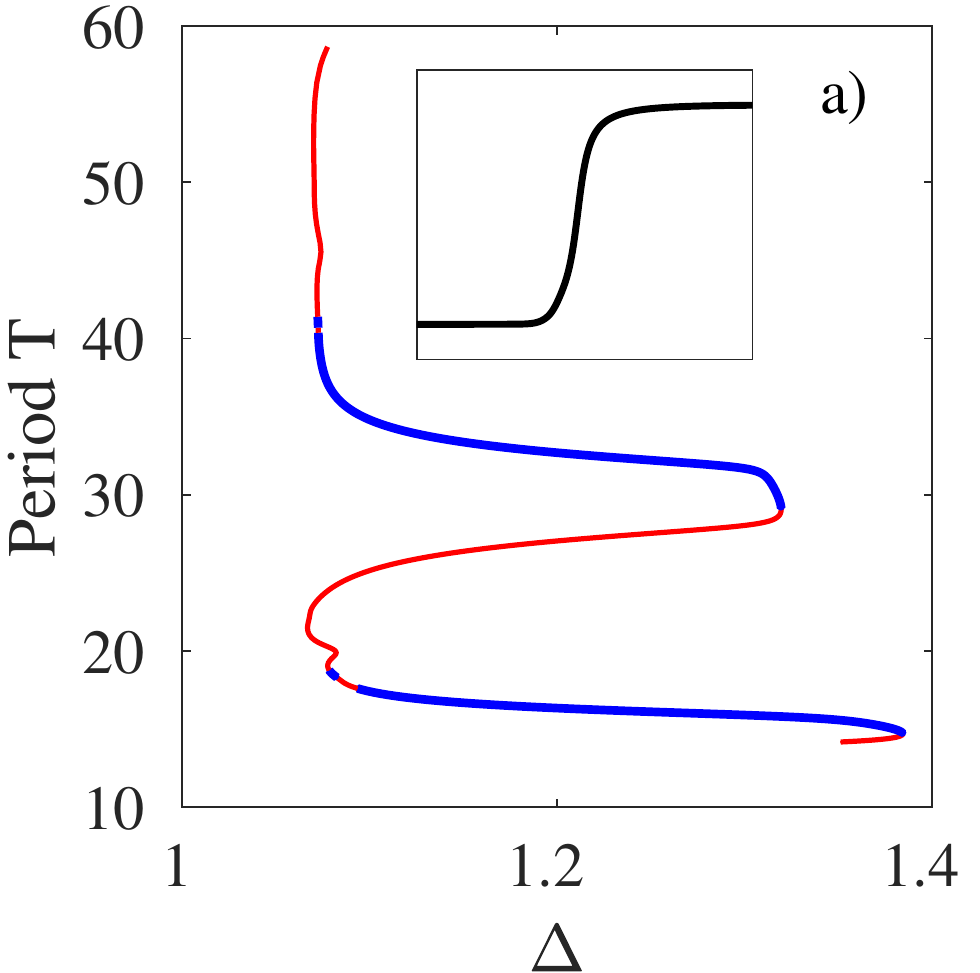}\includegraphics[width=0.49\columnwidth]{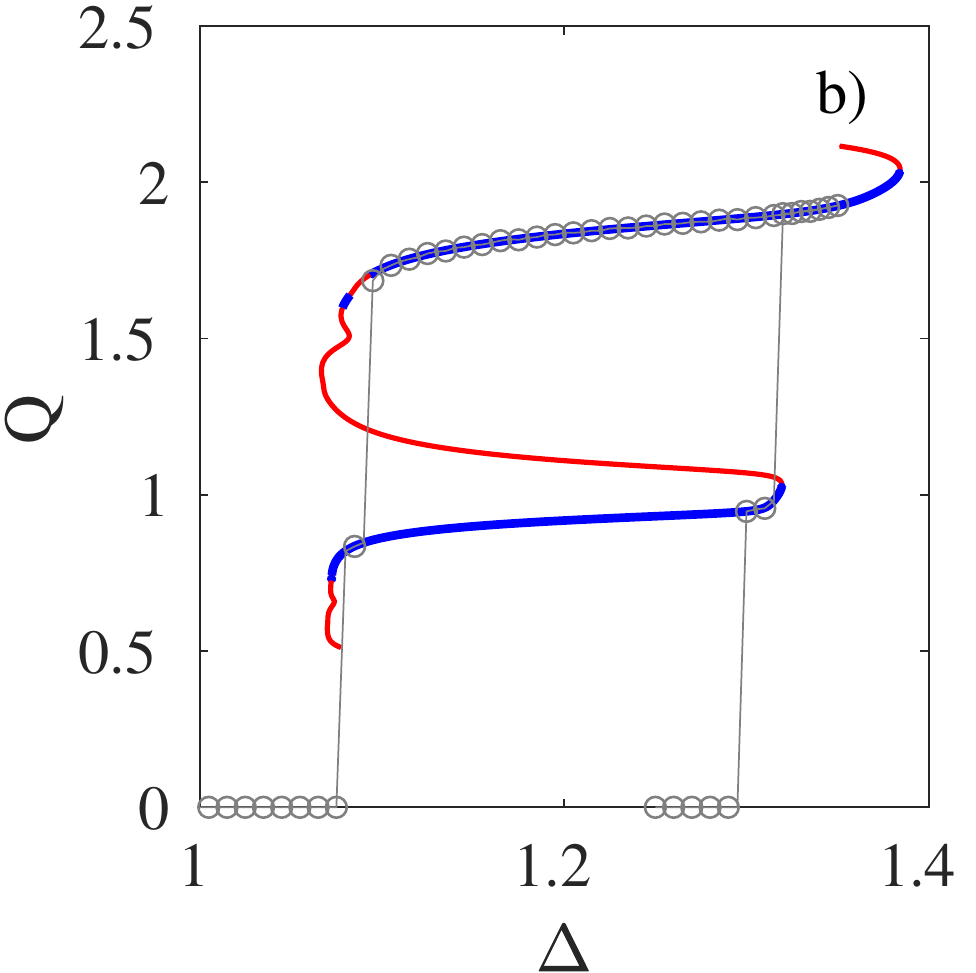}
\caption{Bifurcation diagram of periodic solutions of Eq.~\eqref{eq:adler}.
Here blue bold lines show stable solutions while red thin lines show
unstable ones. In (a) the y-axis displays the period $T$ of the solution
while in (b) the topological charge $Q$ \eqref{eq:topocharge} is
shown on the y-axis. The inset in (a) represents the exemplary periodic
solution profile. The black crosses in (b) show results of a direct
numerical integration with varying detuning $\Delta$. Parameters
are $\chi=0.3$, $\tau=30$ and $\psi=1.4$.}
\label{fig:bifurcationdiagramm} 
\end{figure}

In the region surrounding the SN bifurcations the system is excitable.
A small perturbation from the stable state $\theta_{s}$ decays exponentially
because the system is linearly stable. If however the system is close
to the SN bifurcation, the states $\theta_{s}$ and $\theta_{u}$
are close. In this case, a finite size perturbation can excite the
system beyond the linearly unstable state $\theta_{u}$ leading to
a trajectory towards $\theta_{s}+2\pi$ instead of relaxing back directly
towards $\theta_{s}$. This means that a small perturbation of a stable
state can lead to a large orbit in phase space. If we now introduce
time delay into this excitable system, one excitation at the time
$t-\tau$ can lead to another excitation at time $t$ resulting in
a periodic repetition of excitations with an approximate period of
the delay time $\tau$. In the inset of figure \ref{fig:bifurcationdiagramm}
(a) an example of a periodic orbit from $\theta_{s}$ to $\theta_{s}+2\pi$
is shown. In the following, we shall concentrate our attention on
positive (upward) kinks. Downward (anti-kinks) can be deduced from
the kink regimes by using the symmetry of Eq.~.(\ref{eq:adler})
$\left(\theta,\Delta,\psi\right)\rightarrow-\left(\theta,\Delta,\psi\right)$.
Hence if kinks are found for parameters $\left(\Delta,\chi,\psi\right)=\left(\Delta^{*},\chi^{*},\psi^{*}\right)$,
identical anti-kinks exist at $\left(\Delta,\chi,\psi\right)=\left(-\Delta^{*},\chi^{*},-\psi^{*}\right)$.

We further investigate these regenerative excitable orbits with the
help of \texttt{dde-biftool}. They are implemented as $T-$periodic
orbits, going from from $\theta_{s}$ towards $\theta_{s}\;\left[2\pi\right]$.
We note that $\left[2\pi\right]$ modulo operator is implemented automatically
in the recent versions of \texttt{dde-biftool}. The periodic solutions
were then continued in $\left(\Delta,\chi,\psi\right)$ and $T$ while
adjusting the profile of $\theta\left(t\right)$ on an adaptive grid.
The resulting branch of solutions is shown in Fig. \ref{fig:bifurcationdiagramm}(a).
One can observe two stable regions, one having a period slightly above
$\tau=30$ and the other one having a period slightly above $\frac{\tau}{2}=15$.
To achieve a better visualization of the periodic solutions, the topological
charge $Q$, corresponding to the number of $2\pi$-phase differences
per time delay $\tau$ is introduced:
\begin{align}
Q= & \frac{\tau}{T_{1}-T_{0}}\int_{T_{0}}^{T_{1}}\frac{\text{d}\theta}{2\pi}.\label{eq:topocharge}
\end{align}

For the parameters used in Fig. \ref{fig:bifurcationdiagramm} ($\psi=1.4$,$\chi=0.3$),
the steady state $\theta_{s}$ vanishes at $\Delta=1.296$ but the
continuation clearly shows stable periodic solutions for values of
$\Delta>1.296$. To investigate this region further, a direct numerical
integration was performed starting with the state $\theta_{s}$ at
$\Delta=1.25$. After integrating for a time long enough to ensure
stability of the result, the parameter $\Delta$ was changed and the
next step of integration was performed starting with the result of
the former integration. First $\Delta$ was increased up to $1.35$
which lies outside of the range of existence of $\theta_{s}$. After
reaching $\Delta=1.35$ the process was continued in the other direction
till $\Delta=1$. For each step the topological charge $Q$ was calculated
using Eq. \eqref{eq:topocharge}. In Fig. \ref{fig:bifurcationdiagramm}(b)
the topological charge $Q$ is shown for the results of the time integration
and the continuation. One can observe that the system starts and stays
in the state $\theta_{s}$ which has a topological charge of $0$
until around $\Delta=1.296$ the system jumps to the stable branch
with $Q\approx1$. At the point where the stable periodic solution
with $Q\approx1$ disappears in a SN bifurcation the system jumps
to the stable solution with $Q\approx2$ which is stable up to values
of $\Delta>1.35$. In the reverse direction where $\Delta$ is decreased
from $\Delta=1.35$ to $\Delta=1$ one can observe a hysteresis because
there is a region ranging from $\Delta\approx1.1$ to $\Delta\approx1.3$
where all three solutions with $Q=0$, $Q\approx1$ and $Q\approx2$
are stable. After reaching $\Delta\approx1.1$ the system falls back
to the solution with $Q\approx1$ which also gets unstable for $\Delta\approx1.075$
resulting in the system falling back to $\theta_{s}$ with $Q=0$.

\begin{figure*}[!t]
\includegraphics[viewport=20bp 30bp 580bp 570bp,clip,width=2\columnwidth]{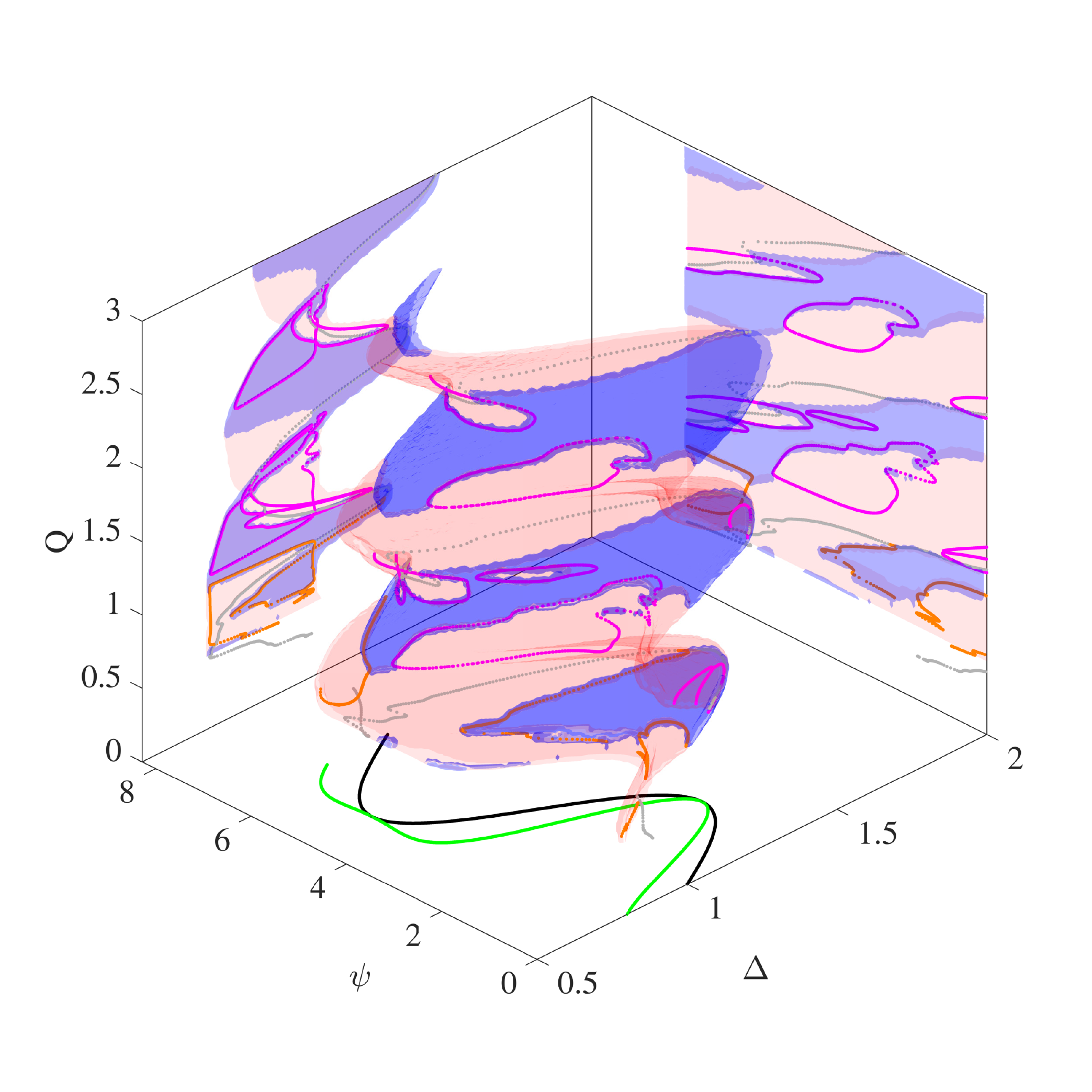}
\caption{Three-dimensional representation of the bifurcation structure of periodic
solutions. Blue and red surfaces represent stable and unstable solutions.
Grey, orange and magenta lines show borders of fold, period doubling
and torus instabilities respectively. The long delay instabilities
of the steady states are included at $Q=0$ (black and green lines)
as a reference. In the background a two-dimensional projection on
to the ($Q$-$\Delta)$ and ($Q$-$\psi)$ planes is shown. Parameters
are $\chi=0.3$ and $\tau=30$.}
\label{fig:3Dview} 
\end{figure*}

A global way of displaying the instabilities of the periodic solutions
such as torus and period-doubling bifurcation consists in representing
the charge $Q$ as a function of $\Delta$ and $\psi$. To this aim,
several branches of periodic solutions were calculated for equidistant
values of $\psi$. The resulting point cloud was then interpolated
into a two-dimensional surface of periodic solutions. There is certainly
a loss of accuracy in this interpolation but the resulting surface
is only used for illustrative purposes. In Fig. \ref{fig:3Dview}
one perspective of this surface is plotted in the three-dimensional
representation, as well as the background instabilities of the steady
state at $Q=0$ and the three types of instabilities for the periodic
solutions. A video of the full thee-dimensional structure can be found
in the Supplementary Material. Here one can clearly see that the stable
regions of periodic solutions form one connected surface that increase
in $Q$ by $1$ if one increases $\phi$ gradually by $2\pi$. This
band of stable solutions is in some cases interrupted by the torus
or period doubling bifurcations leading to unstable regions inside
the stable surface. Since all branches of periodic solutions that
were continued in $\Delta$ for a specific value of $\psi$ represent
a cut of this surface, those regions of instability lead to a splitting
of the stable regions as seen in Fig.~\ref{fig:bifurcationdiagramm}.

\begin{figure}[ht!]
\includegraphics[width=0.8\columnwidth]{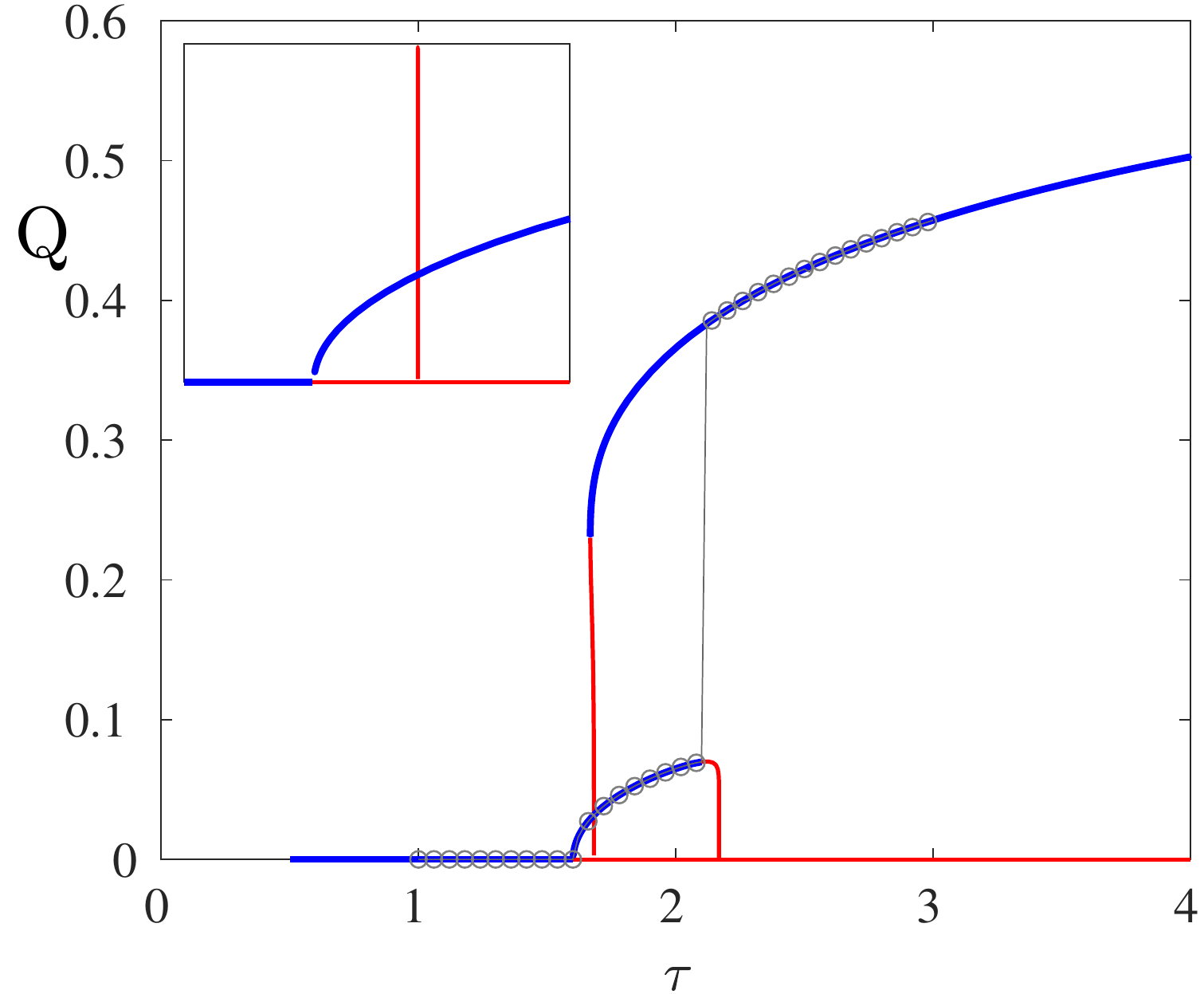}
\caption{Bifurcation diagram in ($Q$-$\tau)$ plane, showing branches of rotating
solutions and librations around the steady state $\theta_{s}$. Unstable
and stable solutions are indicated as red thin lines and blue bold
lines, respectively. Results obtained by direct numerical integration
with varying delay $\tau$ are also shown as a gray circles. Parameters
are $\left(\Delta,\chi,\psi\right)=\left(0.7,1,2.84\right)$.}
\label{fig:globalbifurcation} 
\end{figure}

An interesting kind of periodic solution are those, whose period $T$
are diverging and that can be much greater than the delay $\tau$.
These solutions have a small value of $Q$ and therefore approach
the steady state for which $Q=0$. In Fig.~\ref{fig:3Dview} these
solutions were left out due to computation time. We investigate those
solutions with low topological charge $Q$ to learn more about their
connection to the branch of periodic solutions and to the steady state.
One possible connection involves the AH instability of the steady
state $\theta_{s}$. In Fig.~\ref{fig:hopfcontinuation} we indeed
showed that several AH instabilities appear increasing the time delay
$\tau$. We could therefore prepare the system in a parameter regime
such that there is a stable steady state $\theta_{s}$ for small delay
values and make it AH unstable by increasing the delay time. With
a further increase in $\tau$ the amplitude of the libration grows
up to the point that it induces the nucleation of as a fully developed
rotating solution. Notice that these solutions can not be considered
as LSs since the latter need a stable background. Since an AH instability
leads to small amplitude periodic oscillations and due to the frequent
use of the term \emph{periodic solutions} for the LSs, we will constrict
the use of the term \emph{rotating solutions} to LSs and, more generally,
to the unlocked solutions where the temporal variations of the phase
are unbounded. We will refer to the small oscillations arising from
the AH instabilities as \emph{librations}.

In order to compare the branches corresponding to rotations and librations
in a single bifurcation diagram, we have to adjust our measure $Q$
since using the definition given in Eq.~\eqref{eq:topocharge} both
the steady state and the librations would have a topological charge
of $0$. While this accurately represents the number of $2\pi$ phase
flips that happen in a period $T$, we want to observe the transition
between those states. For this reason a different measure was used
to display the different branches in Fig.~\ref{fig:globalbifurcation}:

\begin{eqnarray}
Q & = & \frac{\delta\theta}{2\pi}\frac{\tau}{T}\label{eq:Charge_bis}
\end{eqnarray}
defining $\delta\theta$ as the amplitude of the phase variation over
the period $T$. In Fig.~\ref{fig:globalbifurcation} the three branches
of solutions are displayed with this new measure on the vertical axis.
One can clearly see the stable oscillating branch emerging from the
AH bifurcation point of the steady state at $\tau_{AH}\approx1.6$.
The branch of rotating solutions, however, does not seem to emerge
from the branch of librations, but to connect with the homogeneous
solution using this measure. One notices that the branch of rotating
solutions reaches values of $Q$ that are much lower than unity. With
our definition of $Q$ given in Eq.~\ref{eq:Charge_bis}, this is
only possible if the period $T$ of the branch of LSs diverges, thereby
indicating a global bifurcation connecting the rotating solution with
the steady states.
\begin{figure}[ht!]
\includegraphics[viewport=20bp 0bp 550bp 244bp,clip,width=1\columnwidth]{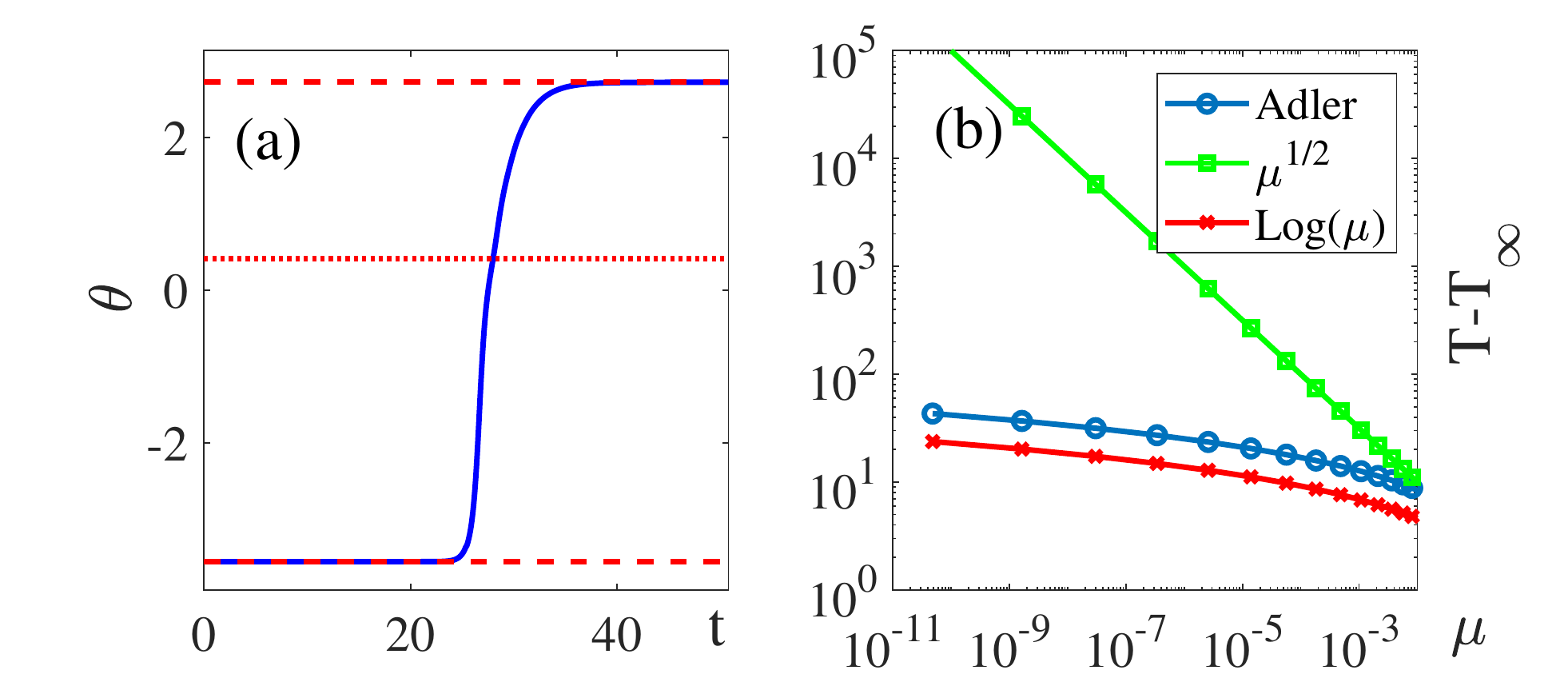}
\caption{(a) Profile of the periodic solution (a) at $\tau=\tau^{*}$ that
connects the unstable steady state $\theta_{u}$ (dashed lines) with
itself. The solution $\theta_{s}$, that is AH unstable at this value
of $\tau$ is marked by a dotted line. (b) Scaling of the period in
the vicinity of the bifurcation point as a function of $\mu=\tau-\tau^{*}$
with $\tau^{*}\simeq1.684511$. Parameters are $\left(\Delta,\chi,\psi\right)=\left(0.7,1,2.84\right)$.}
\label{fig:global_scaling}
\end{figure}

A possible candidate would be a homoclinic bifurcation, however, the
Adler equation without delay is sometimes considered to be the normal
form of the global saddle-node infinite period bifurcation (SNIPER)
\citep{Strogatz2015}. As such, one may expect the global bifurcation
connecting the steady state and the periodic solution of the delayed
Adler equation to be a SNIPER as well. The two possible bifurcations
can be distinguished by the scaling of the period as a function of
the distance $\mu$ to the bifurcation point, while approaching the
latter. Figure~\ref{fig:global_scaling}(a) depicts a profile obtained
as close as possible to the bifurcation point, using $150$ collocation
points and sixth order polynomials, that we approximate at $\tau^{*}\simeq1.684511$,
and for which the period is $T\sim51$. One notices clearly that this
rotating orbit connects the unstable steady state $\theta_{u}$ onto
itself. Figure~\ref{fig:global_scaling}(b) allows us to verify that
the period scales with $\log\mu$ with $\mu=\left|\tau-\tau^{*}\right|$
indicating a homoclinic orbit. We noticed that the transition layer
in Figure~\ref{fig:global_scaling}(a) is relatively smooth indicating
that, considering the low values of $\tau$ used, $\theta\left(t-\tau\right)$
could be expanded in Taylor series. At second order in the truncation,
Eq.~\ref{eq:adler} transforms into an ordinary differential equation
for a forced, damped, nonlinear oscillator whose inertia and damping
terms depend on $\tau$. It is in principle possible to search for
specific values of $\tau$ at which infinite period solutions exist,
leading to an approximation of $\tau^{*}$.

\begin{figure}[ht!]
\includegraphics[width=0.49\columnwidth]{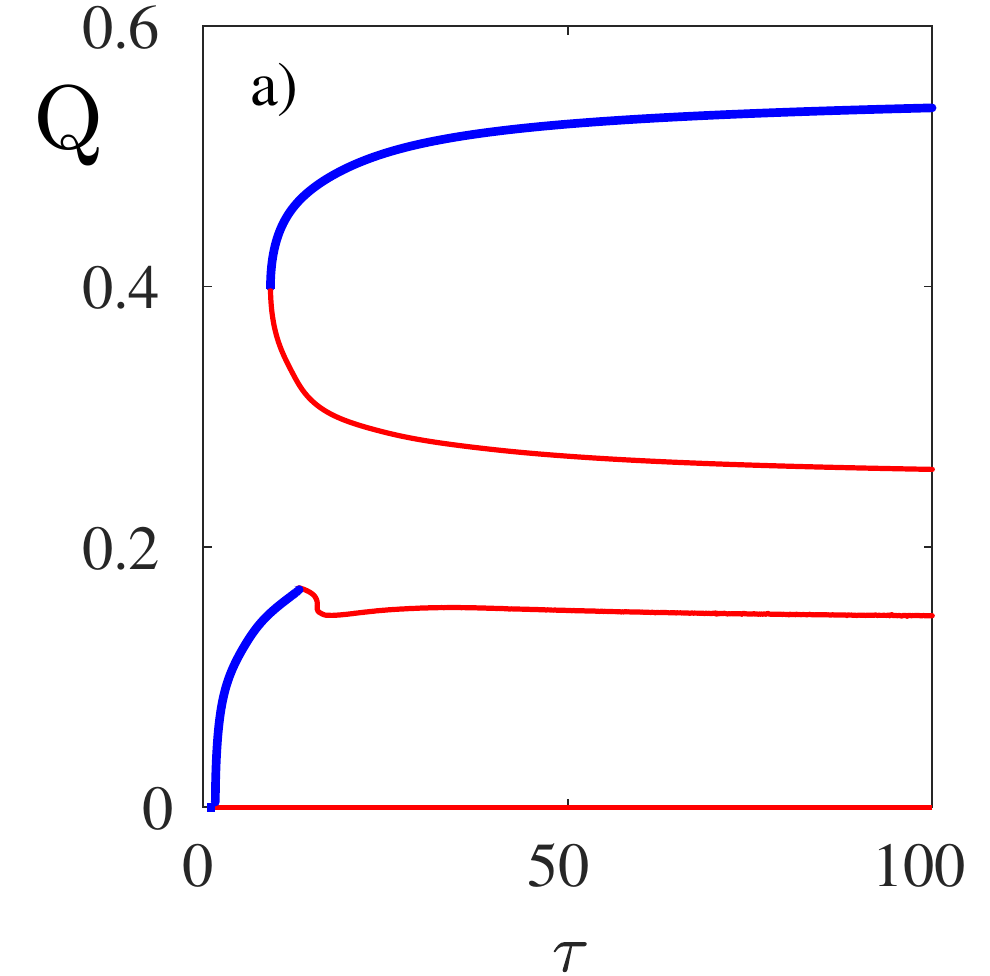}\includegraphics[viewport=0bp 0bp 350bp 290bp,clip,width=0.49\columnwidth,height=4.1cm]{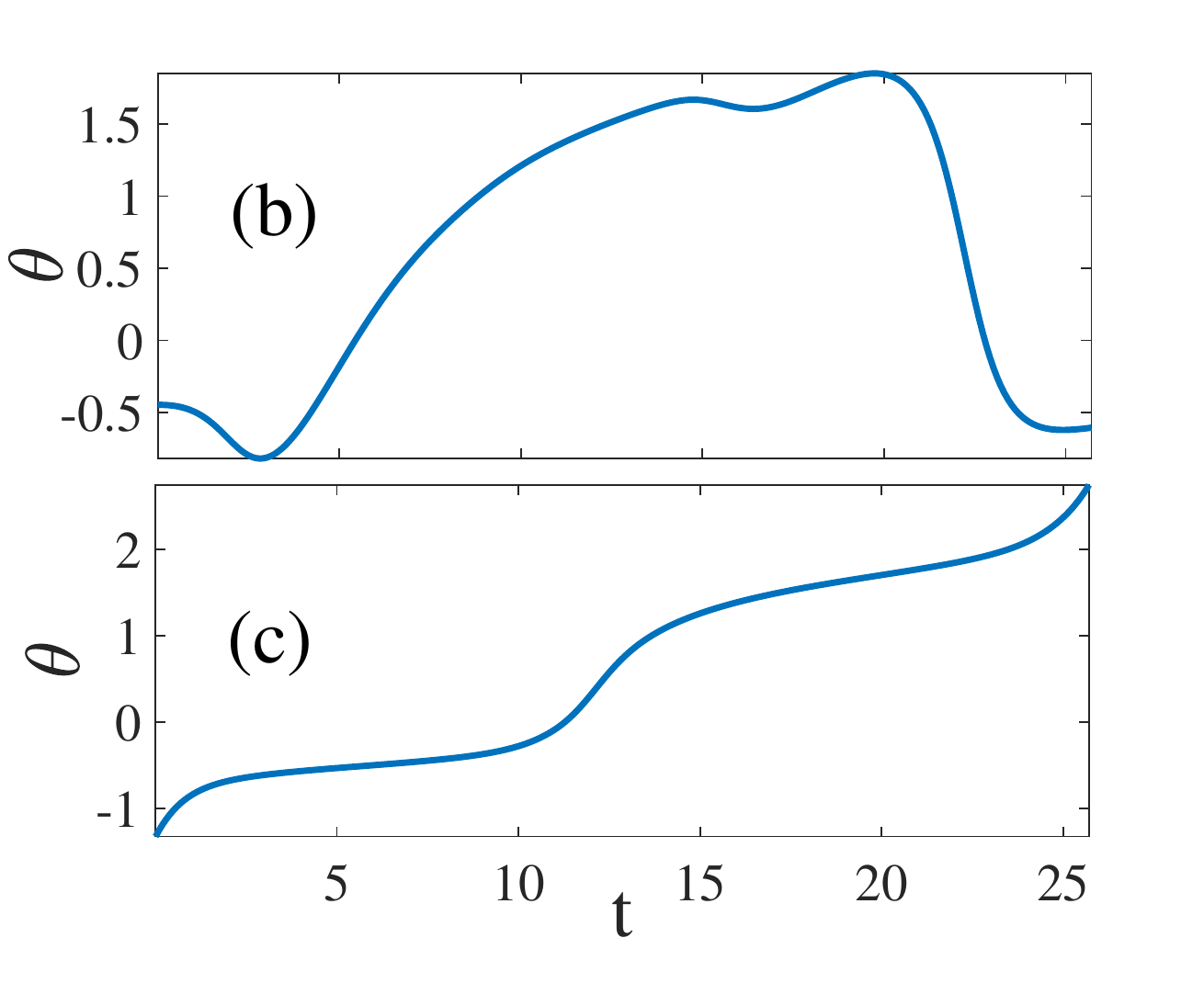}
\caption{(a) Bifurcation diagram showing the branches of libration and of rotation
and the unstable steady branch $\theta_{s}$ as a function of $\tau$.
Parameters are $\left(\Delta,\chi,\psi\right)=\left(0.1,1,2.84\right)$
and the stability is indicated as red thin lines for unstable solutions
and blue bold lines for stable solutions. (b) libration and (c) rotation
are bistable at $\tau=12.87$ and both solutions have a period close
to $2\tau$. }
\label{fig:P2story} 
\end{figure}

\begin{figure}[ht!]
\includegraphics[width=1\columnwidth]{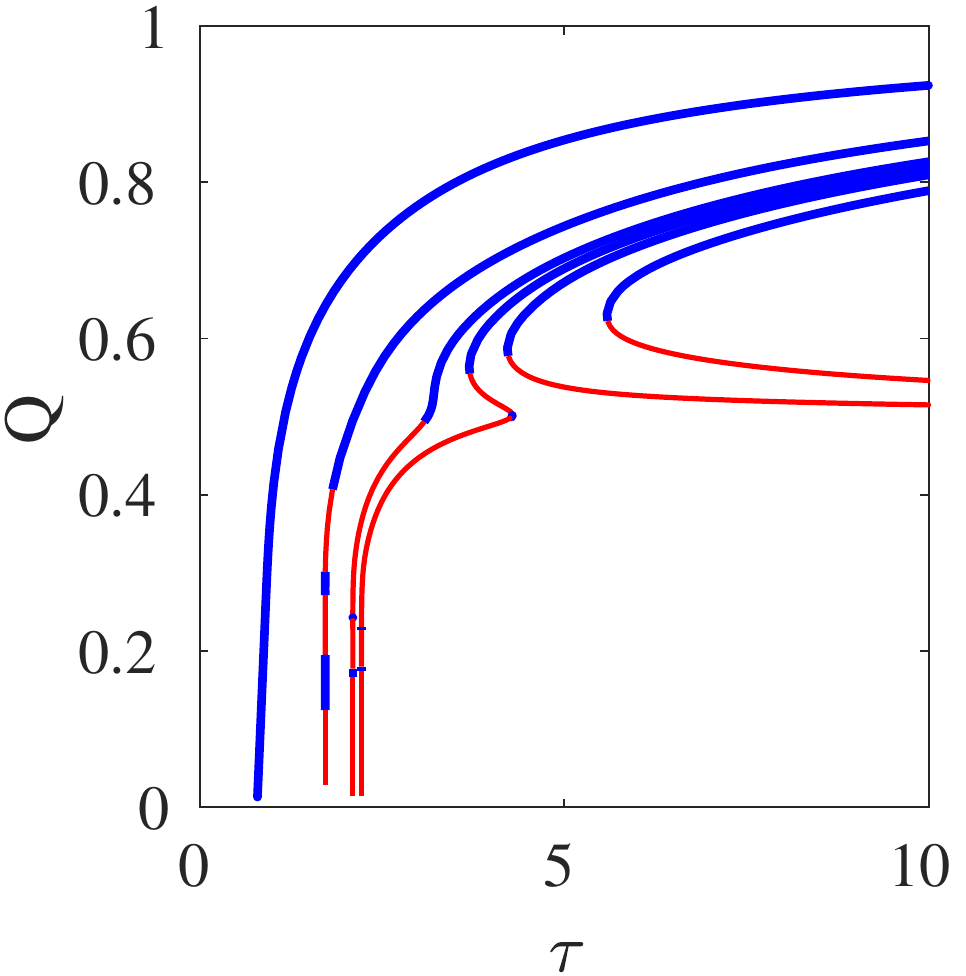} \caption{Bifurcation diagram showing the deformation of the branches of rotations
with varying distance from the saddle node of steady states in a range
of detuning $\Delta\in\left[0.7,2\right]$, while other parameters
are $\chi=1$ and $\psi=\pi/2$. The leftmost branches correspond
to the larger values of $\Delta$ and the stability is indicated as
red thin lines for unstable solutions and blue bold lines for stable
solution. In all cases the steady solution $\theta_{s}$ is stable
and the stable rotating orbits correspond to LSs, for large values
of $\tau$. }
\label{fig:reconnect}
\end{figure}

\begin{figure}[ht!]
\includegraphics[viewport=0bp 0bp 370bp 211bp,clip,width=1\columnwidth]{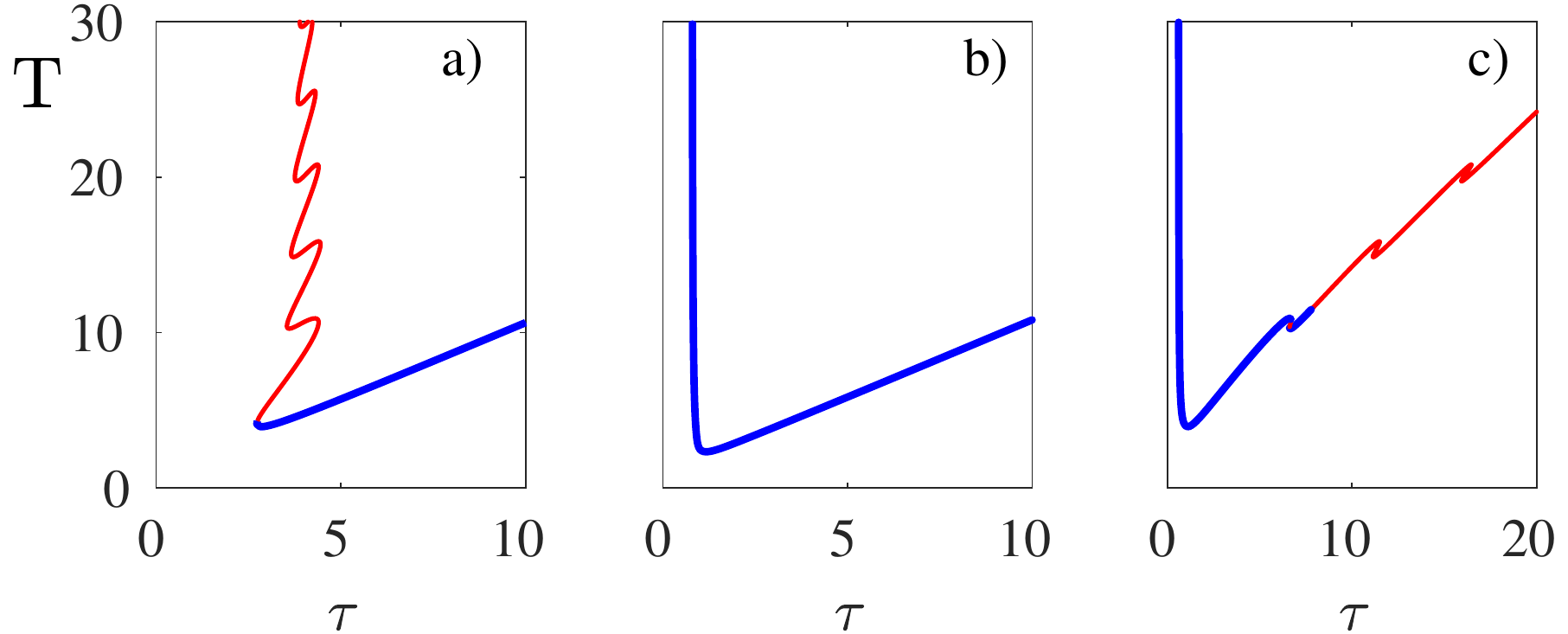}
\caption{Branch of periodic solutions with $\tau$ as a free parameter for
varying feedback phases calculated at the saddle node of steady states
$\Delta=\Delta_{sn}^{+}$. (a) $\left(\Delta,\chi,\psi\right)=\left(1,1,0\right)$
shows snaking on the diverging branch. (b) $\left(\Delta,\chi,\psi\right)=\left(2,1,\pi/2\right)$
shows no snaking. In both cases, the period does not scale with $\mu^{-1/2}$
or with $\log\mu$.}
\label{fig:branchdiverge} 
\end{figure}

More surprisingly, we have found that the homoclinic bifurcation scenario
is not always the reason behind the emergence of rotating solutions.
This can be observed in Fig.~\ref{fig:P2story}(a) where we depict
the bifurcation diagram including steady states, librations and rotating
solutions for a set of parameters $\left(\Delta,\chi,\psi\right)=\left(0.1,1,2.84\right)$
for which the steady states are AH unstable but where the solutions
are far from the SN bifurcation of steady states. The panels Fig.~\ref{fig:P2story}(b,c)
display the profile of these period two (P2) solutions for which $T\sim2\tau$,
which explains why $Q\sim0.5$ for the rotating solution. It is well
known that time delayed systems can give rise to such P2 regimes,
see for instance \citep{N-PRE-04,JAH-PRL-15} and reference therein.

We are interested, in particular, in understanding by which bifurcation
scenario temporal LSs appear. In all the non-pathological cases explored,
i.e. when the background solution is linearly stable, we have always
found the LSs to appear from homoclinic bifurcations, with the period
scaling as $\log\mu$, of from SN bifurcations of periodic solutions.
We represent several branches of rotations for various values of $\Delta$
in Fig.~\ref{fig:reconnect}. Multiple branches of periodic solutions
were calculated for different distances from the SN bifurcation of
the steady states which is at $\Delta_{sn}^{+}=2$ for the given values
of $\psi$ and $\chi$. A transition from a global bifurcation for
the largest values of $\Delta$ towards a SN bifurcation of periodic
solutions is clearly visible. Periodic solutions in the vicinity of
the SN of the steady states show a global bifurcation while for larger
distances of $\Delta$ with respect to $\Delta_{sn}^{+}$ the branch
deforms, while conserving the homoclinic period divergence at the
branching point, and finally shows a regular saddle-node of limit
cycle bifurcation for values of $\Delta<0.825$. The deformation of
the branches in Fig.~\ref{fig:reconnect} indicates that a reconnection
mechanism with other unstable branches of periodic solutions underlies
the qualitative change in the bifurcation scenario. 

Finally, we followed the evolution of the point at which the period
diverges as a function of delay $\tau$ and feedback strength $\mathcal{\chi}$
by calculating several branches \emph{in the limit case} where the
background solution $\theta_{s}$ is marginally stable. The detuning
$\Delta$ was chosen such that the system is on the border of the
steady state SN bifurcation, which is given by $\Delta=\Delta_{sn}^{+}$.
Figure~\ref{fig:branchdiverge} shows the branches for two parameter
sets. In both cases, the period diverges for small delay yet it does
not scales with $\mu^{-1/2}$ or with $\log\mu$. It is particularly
salient in Fig.~\ref{fig:branchdiverge}(b), where the branch shows
a snaking behavior. For larger delays, the solutions get more and
more localized with a period that scales as $T=\tau+r$ with $r=\mathcal{O}\left(1\right)$.
However, the background is marginally stable leading to solutions
that can hardly be called LSs. They correspond to the pathological
case discussed in \citep{YRS-PRL-19} Fig.~4.

\section{effective interaction law }

We now turn our attention to the interaction law governing the relative
motion of multiple LSs that can be embedded in a sufficiently long
value of the time delay. Distant LSs interact via their exponentially
decaying tails. While in many salient examples the solitons are even
functions, as e.g, those found as solutions of the Nonlinear Schrödinger,
Ginzburg-Landau or Lugiato-Lefever equations, the left and right exponential
tails are not necessarily identical in non-parity preserving systems.
As such, the interactions between LSs may not be reciprocal and disobey
with the action-reaction principle, as demonstrated recently in passively
mode-locked laser \citep{JCM-PRL-16}. For our analysis, it is convenient
to factor out the value of the steady state $\theta_{s}$. We define
$\theta\left(t\right)=\theta_{s}+\phi\left(t\right)$ in such a way
that $\phi\left(t\right)$ can represents stable LSs from $0$ to
$2\pi$. Then the right hand side of the original delayed Adler equation
\eqref{eq:adler} transforms to 
\begin{align}
\dot{\phi}= & F\left(\phi,\phi^{\tau}\right)\label{eq:phi}
\end{align}
with 
\begin{eqnarray}
F\left(\phi,\phi^{\tau}\right) & = & \Delta-\sin\left(\theta_{s}+\phi\right)+\chi\sin\left(\phi^{\tau}-\phi-\psi\right)\,,
\end{eqnarray}
where we used the shorthand $\phi^{\tau}=\phi\left(t-\tau\right)$.
Sufficiently far from the kink, the LSs can be approximated by their
exponential tails. The latter govern the approach of the steady state.
The left and right tails can be expanded as 
\begin{align}
\phi(t)= & \phi(-\infty)+\sum_{i}a_{-}^{\left(i\right)}\exp\left(\sigma_{-}^{\left(i\right)}t\right),\label{eq:genexpminus}\\
\phi(t)= & \phi(+\infty)+\sum_{i}a_{+}^{\left(i\right)}\exp\left(\sigma_{+}^{\left(i\right)}t\right)\label{eq:genexpplus}
\end{align}
with the asymptotic values $\phi(-\infty)=0$ and $\phi(+\infty)=2\pi$
and $\sigma_{\pm}^{\left(i\right)}$ being complex eigenvalues and
$a_{\pm}^{\left(i\right)}$ the corresponding eigenvectors. Note that
a real-valued $\sigma_{\pm}^{\left(i\right)}$ leads to a monotonic
tail while $\sigma_{\pm}^{\left(i\right)}\in\mathbb{C}$ induces oscillatory
tails. For the solution to remain bounded, all coefficients $a_{+}^{\left(i\right)}$
with $\sigma_{+}^{\left(i\right)}>0$ and $a_{-}^{\left(i\right)}$
with $\sigma_{-}^{\left(i\right)}<0$ have to be zero. If the LSs
are sufficiently far away from each other, their interaction is governed
by the slowest decaying mode; we denote $\sigma_{+}$ (resp. $\sigma_{-}$)
the eigenvalues with smallest negative (resp. positive) real part
associated with the eigenvector $a_{\pm}$. As the eigenvalue can
be complex, the tail is in general approximated by
\begin{align}
\phi(t) & =\phi(\pm\infty)+\Re\left(a_{\pm}e^{\sigma_{\pm}t}\right)\,.\label{eq:exponentialLS}
\end{align}

Because we are considering periodic solutions with period $T=\tau+r$,
where $r$ is defined as the solution drifts $r=T-\tau$, we can rewrite
Eq. \eqref{eq:phi} as an advanced time-delayed equation \citep{N-PRE-04,YRS-PRL-19},
replacing $\phi\left(t-\tau\right)$ by $\phi\left(t+r\right)$. How
$T-$periodic solutions approach the uniform state is found by inserting
the exponential expansion Eqs.~(\eqref{eq:genexpminus},\eqref{eq:genexpplus})
into the advanced delayed equation $\dot{\phi}=F\left(\phi,\phi^{-r}\right)$.
A linear analysis allows to obtain the equation governed the eigenvalues
$\sigma_{\pm}^{\left(i\right)}$ as
\begin{align}
\sigma_{\pm}^{\left(i\right)}= & A+B\exp\left(\sigma_{\pm}^{\left(i\right)}r\right)\,,
\end{align}
where $A$ and $B$ are the same coefficients as in Eqs. \eqref{eq:A}
and \eqref{eq:B}. The exponents of the expansion are therefore the
solutions of the eigenvalue problem for the steady states, in which
the delay $\tau$ is replaced with a small negative time $r$. Notice
that $r$ can only be obtained numerically using direct time integration
or during continuation with \texttt{dde-biftool} while the amplitudes
$a_{\pm}$ are obtained by a best fit of the tails of the LS using
Eq.~(\ref{eq:exponentialLS}).

To approximate the interaction between two distant LSs located in
$x_{1}$ and $x_{2}$ with $x_{1}<x_{2}$, we assume an ansatz 
\begin{eqnarray}
\phi(t) & = & \phi_{1}(t)+\phi_{2}\left(t\right),\label{eq:multiLS}
\end{eqnarray}
where $\phi_{i}(t)=\Phi\left[t-x_{i}(t)\right]$ is a $2\pi$ kink
centered around the position $x_{i}\left(t\right)$ and the function
$\Phi$ is a $T-$periodic solution that verifies the equation of
motion of a single LS, i.e. $\dot{\Phi}=F\left(\Phi,\Phi^{\tau}\right)$.
One can expand the left hand side of Eq.~\eqref{eq:phi} and find
\begin{eqnarray}
\dot{\phi} & = & \left(1-\dot{x}_{1}\right)\dot{\phi}_{1}-\left(1-\dot{x}_{2}\right)\dot{\phi}_{2}\,.
\end{eqnarray}

In the vicinity of the second LS one can approximate the right hand
side of Eq.~\eqref{eq:phi} as
\begin{eqnarray*}
F\left(\phi,\phi^{\tau}\right) & \simeq & F\left(\phi_{2},\phi_{2}^{\tau}\right)+\phi_{1}\frac{\partial F}{\partial\phi}\left(\phi_{2},\phi_{2}^{\tau}\right)+\phi_{1}^{\tau}\frac{\partial F}{\partial\phi^{\tau}}\left(\phi_{2},\phi_{2}^{\tau}\right).
\end{eqnarray*}
Expressing the instantaneous and delayed tail of $\phi_{1}$ as 
\begin{eqnarray}
\phi_{1} & = & \phi\left(t-x_{1}\right)=\Re\left[a_{+}e^{\sigma_{+}\left(t-x_{1}\right)}\right],\\
\phi_{1}^{\tau} & = & \phi\left(t+r-x_{1}^{\tau}\right)=\Re\left[a_{+}e^{\sigma_{+}\left(t+r-x_{1}^{\tau}\right)}\right],
\end{eqnarray}
one can derive the following equation: 
\begin{eqnarray}
-\dot{x}_{1}\dot{\phi}_{1}-\dot{x}_{2}\dot{\phi}_{2} & = & \Re\left\{ a_{+}e^{\sigma_{+}\left(t-x_{1}\right)}\left[\partial_{1}F\left(\phi_{2},\phi_{2}^{\tau}\right)-\sigma_{+}\right]\right.\nonumber \\
 & + & \left.a_{+}e^{\sigma_{+}\left(t+r-x_{1}^{\tau}\right)}\partial_{2}F\left(\phi_{2},\phi_{2}^{\tau}\right)\right\} \,.\label{eq:rule}
\end{eqnarray}

Before proceeding to the projection of Eq.~(\ref{eq:rule}) onto
the neutral mode of the adjoint problem, some useful simplifications
can be performed by noticing that the displacements $\dot{x}_{i}$
are already small quantities, as their source stem from overlap integrals.
As such, in the equation for $\dot{x}_{2}$, the cross inertia term
$\dot{x}_{1}$ will be multiplied by an overlap integral $\int\dot{\phi}_{2}^{\dagger}\dot{\phi}_{1}$
which is a small quantity. Similarly, the time delay $x_{1}\left(t-\tau\right)$
in Eq.~(\ref{eq:rule}) can be expanded to first order in $\tau$,
which will generate another contribution proportional to $\dot{x}_{1}$,
that can be neglected for the same reason.

For the $T-$periodic solution $\Phi\left(t\right)$, the Floquet
analysis is obtained setting $\phi\left(t\right)=\Phi\left(t\right)+\varepsilon u\left(t\right)$
which results in a linear delay equation with $T$-periodic coefficients
\begin{eqnarray}
\dot{u} & = & a\left(t\right)u\left(t\right)+b\left(t\right)u\left(t-\tau\right)\,.\label{eq:Direct}
\end{eqnarray}

Performing the linear stability analysis of Eq.~\ref{eq:Direct}
yields a neutral eigenfunction with the Floquet multiplier $\mu=1$
and denoted $u_{0}=\dot{\Phi}$ that represents a translation along
the periodic orbit. The adjoint problem is defined with respect to
the standard scalar product of two functions $\left(u,v\right)=\int u\bar{v}dt$
and reads
\begin{eqnarray}
-\dot{v} & = & a^{\dagger}\left(t\right)v\left(t\right)+b^{\dagger}\left(t+\tau\right)v\left(t+\tau\right)\label{eq:Adjoint}
\end{eqnarray}
which can be simplified using that $a$ and $b$ are real valued scalar
functions for which $\left(a^{\dagger},b^{\dagger}\right)=\left(a,b\right)$.
Equation (\ref{eq:Adjoint}) corresponds to delayed equation with
a negative delay and $T-$periodic coefficients that must be integrated
backward in time. The adjoint problem given by Eq.~\ref{eq:Adjoint}
possesses the same eigenvalues than the direct problem Eq.~\ref{eq:Direct},
although the eigenvalue $\mu=1$ is associated to a different eigenvector
that we note in the following $v_{0}=\dot{\Phi}^{\dagger}$. We note
the eigenvectors of Eq.~\ref{eq:Direct} and of Eq.~\ref{eq:Adjoint}
$\left\{ u_{j}\right\} $ and $\left\{ v_{j}\right\} $, respectively.
Since these two sets are bi-orthogonal with respect to they other,
we have, by a proper normalization choice, $\left(u_{j,}v_{k}\right)=\delta_{jk}$.
The motion induced by the tail of a distant LS over another can be
calculated by projecting its contribution in Eq.~(\ref{eq:rule})
along $v_{0}$. In order to find this particular eigenfunction, we
partially diagonalized the evolution operator corresponding to Eq.~(\ref{eq:Adjoint})
using the Implicitly Restarted Arnoldi Method (IRAM) \citep{IRAM}
with a method similar to the one described in \citep{GJT-NC-15}.
However, as we are mainly interested in the LS solutions that are
stable, the largest multiplier in both the direct and adjoint linearized
problem is $\mu=1$. Hence one could evaluate $\dot{\Phi}^{\dagger}$
by simply integrating numerically Eq.~(\ref{eq:Adjoint}) starting
from a random initial condition until the solution converges towards
a periodic profile. Repeating the same analysis in the vicinity of
the first LS and using the asymptotic expansion for the second LS
that involves $\left(a_{-},\sigma_{-}\right)$ yields the equation
of motion for $x_{1}$ and $x_{2}$ as
\begin{eqnarray}
\dot{x}_{2} & = & \Re\left[F_{+}e^{\sigma_{+}(x_{2}-x_{1})}\right],\label{eq:rhs+}\\
\dot{x}_{1} & = & \Re\left[F_{-}e^{\sigma_{-}(x_{1}-x_{2})}\right],\label{eq:rhs-}
\end{eqnarray}
with the coefficients $F_{\pm}$ given by:\begin{widetext} 

\begin{align}
F_{\pm}= & \frac{\int_{-\infty}^{\infty}\dot{\Phi}^{\dagger}\left(s\right)e^{\sigma_{\pm}s}\times\left[\frac{\partial F}{\partial\phi}\left(\Phi,\Phi^{\tau}\right)-\sigma_{\pm}+e^{\sigma_{\pm}r}\frac{\partial F}{\partial\phi^{\tau}}\left(\Phi,\Phi^{\tau}\right)\right]\text{d}s}{\int_{-\infty}^{\infty}\dot{\Phi}^{\dagger}\left(s\right)\dot{\Phi}\left(s\right)\text{d}s}a_{\pm}\,.
\end{align}

\end{widetext} 

The equation of motion for the distance between the two LSs $l=x_{2}-x_{1}$
can be recast as to depend on the gradient of a potential $U\left(l\right)$
\begin{eqnarray}
\dot{l} & = & -\frac{dU}{dl}\;,\;U\left(l\right)=\Re\left(\frac{F_{-}}{\sigma_{-}}e^{-\sigma_{-}l}-\frac{F_{+}}{\sigma_{+}}e^{\sigma_{+}l}\right).\label{eq:rhsdist}
\end{eqnarray}

\begin{figure}[ht!]
\includegraphics[width=0.5\columnwidth]{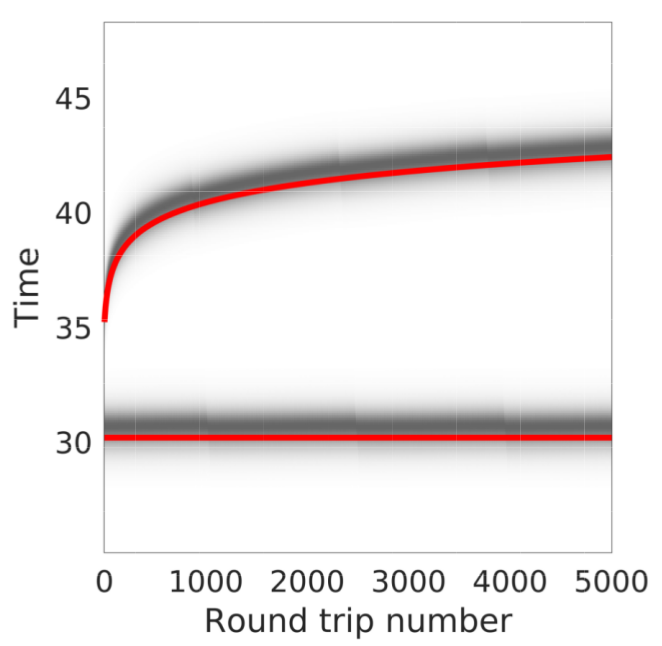}\includegraphics[width=0.5\columnwidth]{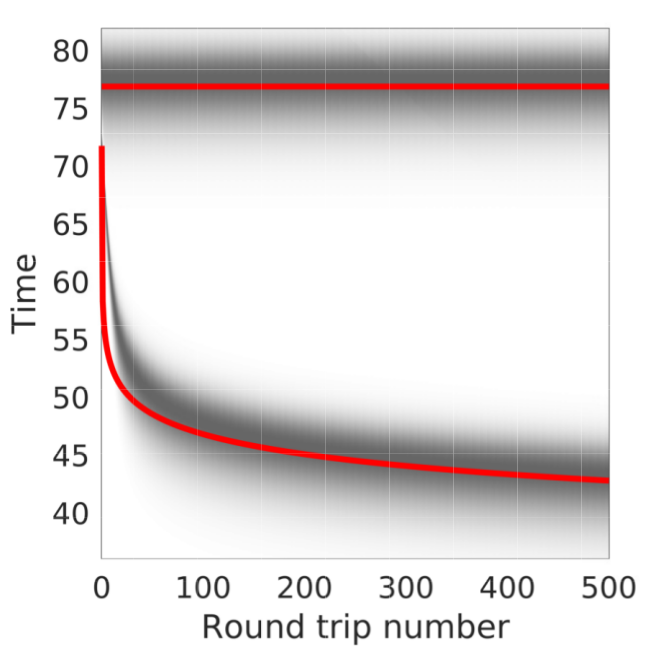}\caption{Asymmetrical interactions between LSs. The DNS of the delayed Adler
equation Eq.~\eqref{eq:adler} and of the reduced model Eqs.~\eqref{eq:rhs+},\eqref{eq:rhs-}
are shown as shades of gray and red lines, respectively. Left: Repulsive
``causal'' interaction for parameters $\left(\Delta,\chi,\psi,\tau\right)=\left(0.9393,0.99,0,100\right)$
leading to $\left(F_{+},F_{-}\right)=\left(4.88,0.003\right)$ and
$\left(\sigma_{+},\sigma_{-}\right)=\left(-0.8,1.21\right)$. Right:
Repulsive ``anti-causal'' interaction for parameters $\left(\Delta,\chi,\psi,\tau\right)=\left(0.9974,2,1.396,100\right)$
leading to $\left(F_{+},F_{-}\right)=\left(5\times10^{-4},-4159\right)$
and $\left(\sigma_{+},\sigma_{-}\right)=\left(-0.48,0.4\right)$.}
\label{fig:interaction} 
\end{figure}

\begin{figure}[ht!]
\includegraphics[width=0.5\columnwidth]{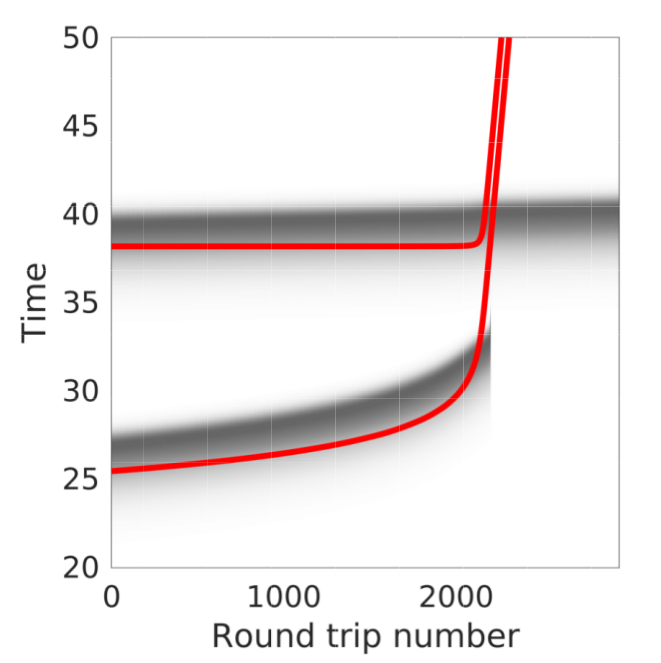}\includegraphics[width=0.5\columnwidth]{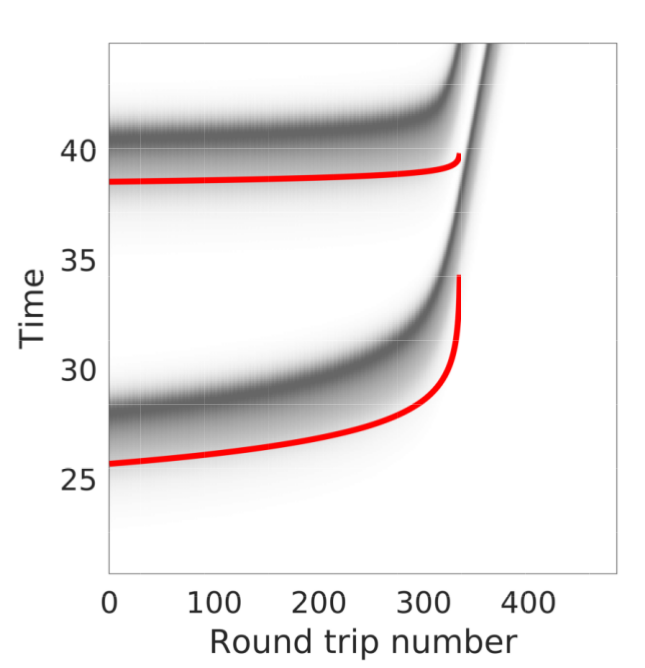}\caption{Creation of drifting bound states of LSs. The DNS of the delayed Adler
Eq.~\eqref{eq:adler} and of the reduced model Eqs.~\eqref{eq:rhs+},\eqref{eq:rhs-}
are shown as shades of gray and red lines, respectively. Left: One
of the LSs in the Adler equation disappear while the effective model
shows the presence of the bound state. Parameters are $\left(\Delta,\chi,\psi,\tau\right)=\left(0.842,0.99,0.698,100\right)$
leading to $\left(F_{+},F_{-}\right)=\left(341,2\right)$ and $\left(\sigma_{+},\sigma_{-}\right)=\left(-1.7,0.6\right)$.
Right: A molecule is formed in the Adler equation while the effective
model give rise to a singularity. Parameters are $\left(\Delta,\chi,\psi,\tau\right)=\left(1.1,0.5,1.323,100\right)$
leading to $\left(F_{+},F_{-}\right)=\left(69,536\right)$ and $\left(\sigma_{+},\sigma_{-}\right)=\left(-0.9,0.92\right)$.}
\label{fig:spaceship}
\end{figure}

In order to visualize the interaction between the LSs we use a pseudo-space-time
representation as introduced in \citep{GP-PRL-96}. This is achieved
by cutting the time trace in slices of one period $T$ and rearranging
them in a second dimension. This is similar to the approach of multiple
timescale analysis, since we display the dynamics happening in one
period on one axis and show the slow evolution from one period to
the next on the other axis. One example of such a pseudo-space time
representation can be seen in Fig.~\ref{fig:interaction} where we
prepared two LSs close to each other as an initial condition. The
vertical axis represents the fast time scale and goes from $0$ to
$T$, yet only the region containing the two LSs is shown. The horizontal
axis displays the round-trip number. The results of the direct numerical
simulations (DNSs) are displayed in gray scale of the variable $\cos\phi$.
In this way, the steady states appear white and the LSs appear in
a shade of gray representing their distance to $\theta_{s}$. The
red lines overlaying the diagram are the results of an integration
of the reduced system given by Eqs.~\eqref{eq:rhs+},\eqref{eq:rhs-}.

One can generally observe a good agreement between the results of
the reduced equations of motion Eqs.~\eqref{eq:rhs+},\eqref{eq:rhs-}
and that of the DNS, as seen for instance in Fig.~\ref{fig:interaction}
when the LSs are not too close. The non-reciprocity of the interactions
is clearly observed and it is mainly due to the asymmetry of the tails
of the LSs and of the overlap integrals; in general $\sigma_{+}\neq-\sigma_{-}$
and $F_{+}\neq-F_{-}$. In particular, $F_{+}>0$ (resp. $F_{+}<0$)
corresponds to LS$_{2}$ being repulsed (resp. attracted) by LS$_{1}$
while and $F_{-}<0$ (resp. and $F_{-}>0$) corresponds to LS$_{1}$
being attracted (resp. repulsed) by LS$_{1}$. Notice that the left
panel situation in Fig.~\ref{fig:interaction} corresponds very well
to the experimental results in Figs.~1(c) of \citep{GJB-CHA-17}.
While these results can be reproduced numerically with a Class-C laser
model in Fig.~5 of \citep{GJB-CHA-17}, we show that the interaction
are properly accounted for in our simplified Adler model.

The absence of parity make it so that it is possible to obtain stable
bound states even in the cases in which $\sigma_{\pm}\in\mathbb{R}$.
We show how non-reciprocal attractive interactions can lead to drifting,
stable bound states in Fig.~\ref{fig:spaceship}. Unfortunately,
as the LSs are extremely close to each other, the reduced set of Eqs.~\eqref{eq:rhs+},\eqref{eq:rhs-}
diverges in some cases instead of showing a molecule bound state at
this parameter set. Yet, other parameter sets allows finding such
drifting bound states. For very small distance between the LSs the
assumption of weak, single exponential long range interaction is not
valid anymore. A possible improvement over Eqs.~\eqref{eq:rhs+},\eqref{eq:rhs-}
would be to include higher order terms in the approximation of the
tails in Eq.~\eqref{eq:exponentialLS} and the projection onto the
weakly damped eigenmodes of Eq.~\ref{eq:Adjoint}.

\begin{figure}[ht!]
\includegraphics[width=0.49\columnwidth]{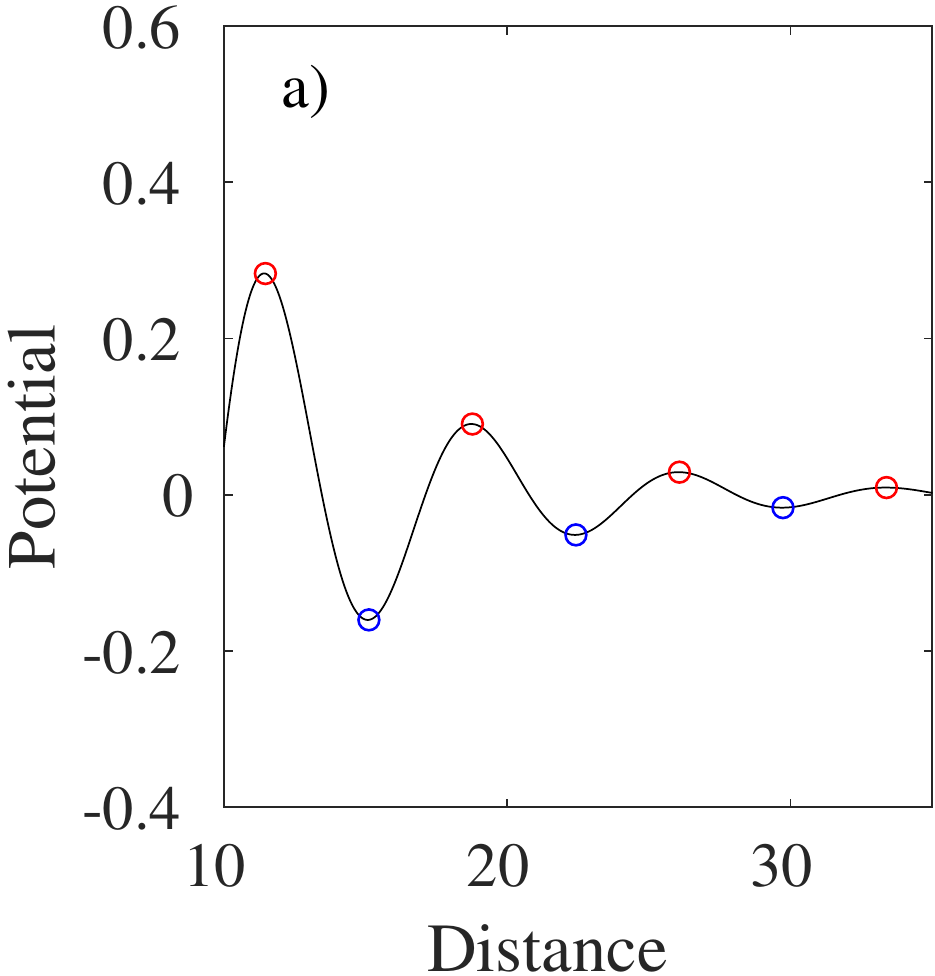}\includegraphics[width=0.49\columnwidth]{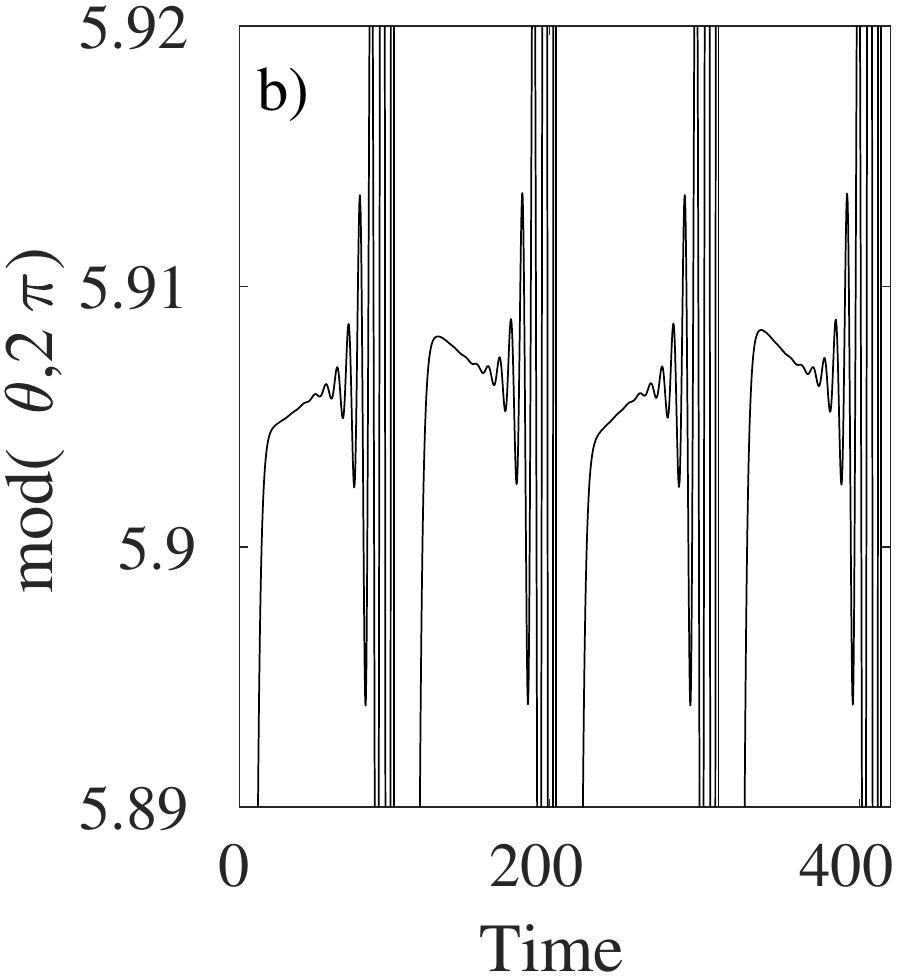}\\
 \includegraphics[width=1\columnwidth]{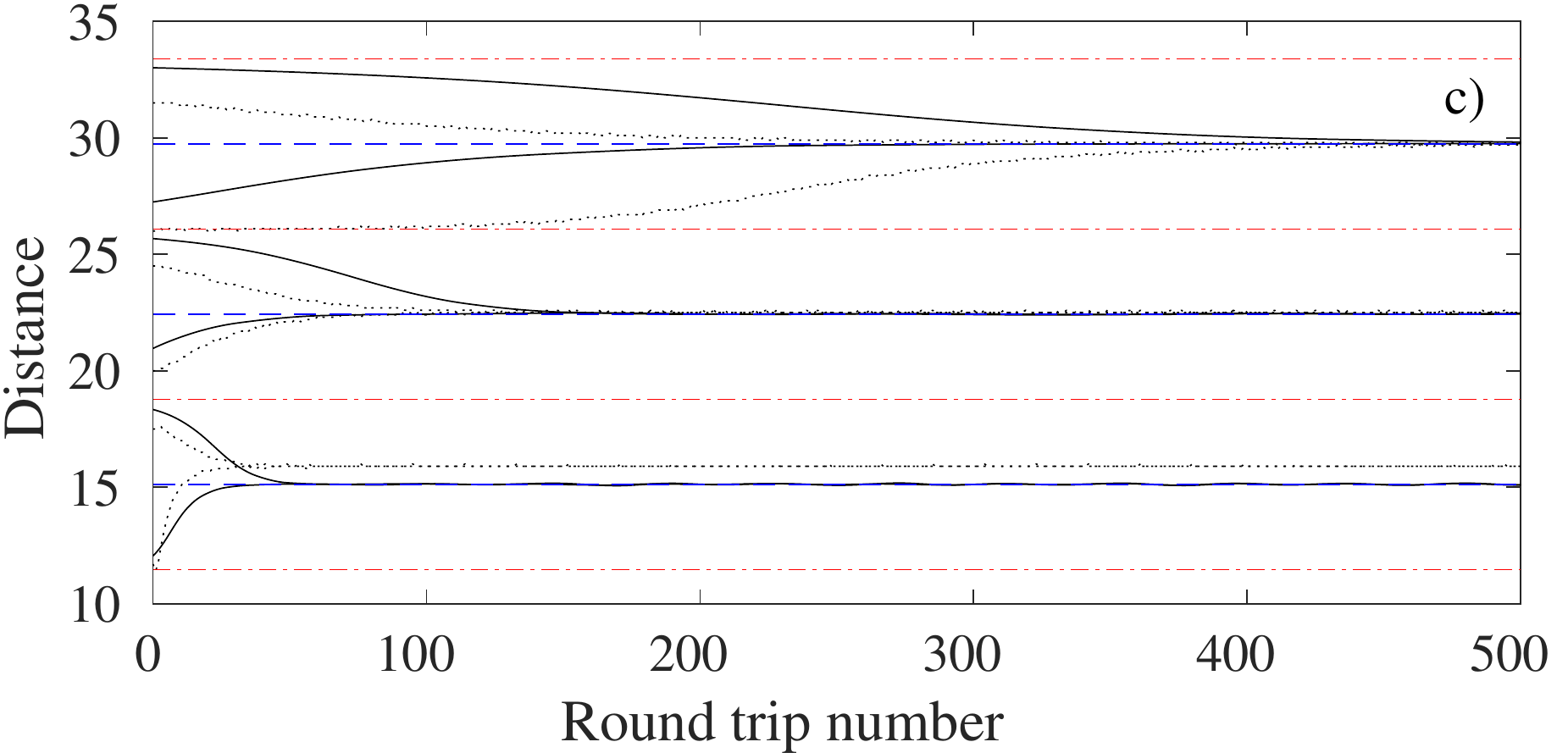}
\caption{(a) Potential of the distance between two LSs. (b) DNS of the Adler
Eq.~\eqref{eq:adler} showing a precursor of a period-two instability
of a single LS. (c) Evolution of the distance between two LSs obtained
from DNS of Adler Eq.~\eqref{eq:adler} in dotted black lines and
reduced model Eq.~\eqref{eq:rhs+},\eqref{eq:rhs-} in solid black
lines as well as minima and maxima of the potential in blue dashed
and red dot-dashed lines. Parameters are $\left(\Delta,\chi,\psi,\tau\right)=\left(0.49,0.99,2.1,100\right)$
in all three cases.}
\label{fig:locking} 
\end{figure}

There is a special regime in parameter space, where the exponential
tails become complex. This region lies in the vicinity of the AH instability
of the steady states and the tails of the LS starts to oscillate at
the frequencies given by $\Im\left(\sigma_{\pm}\right)$. This opens
the possibility for multiple roots for Eq.~\eqref{eq:rhsdist} and
thus a potential $U\left(l\right)$ with multiples (almost) equidistant
minima and maxima. One set of parameters $\left(\Delta,\chi,\psi,\tau\right)=\left(0.49,0.99,2.1,100\right)$
exhibiting this behavior is displayed in Fig.~\ref{fig:locking}.
The corresponding potential $U\left(l\right)$ is depicted in Fig.~\ref{fig:locking}(a).
The DNS of the Adler Eq. \eqref{eq:adler} with two LSs at varying
distances was performed and Fig.~\ref{fig:locking}(c) shows the
evolution of the distance between the LSs over multiple round-trips.
The maxima and minima of the potential are shown in red and blue dashed
lines, respectively while the distances obtained from the reduced
Eqs.~\eqref{eq:rhs+}\eqref{eq:rhs-} are displayed in black lines.
The distance obtained from DNS is also shown in dotted lines. The
results of the DNS do not seem to match very well with that of the
reduced model in this particular case. While converging to the same
distances after many round-trips the transients are different. The
reason for the discrepancy was found in the pathological values of
the parameters we chose for the periodic solution; The feedback phases
$\psi>\pi/2$ bring the system close to a period doubling regime.
A precursor of that period doubling instability can be observed in
the DNS of a single LS which is shown in Fig.~\ref{fig:locking}(b).
The time trace is zoomed into the vicinity of the steady state $\theta_{s}$
to show the small deviation between the relaxations into the steady
state happening every other period. It is well-known that period-two
oscillations of LSs can modify strongly their interactions, up to
the point of canceling the coarsening dynamics as shown in \citep{JAH-PRL-15}.

\begin{figure}[ht!]
\includegraphics[width=0.8\columnwidth]{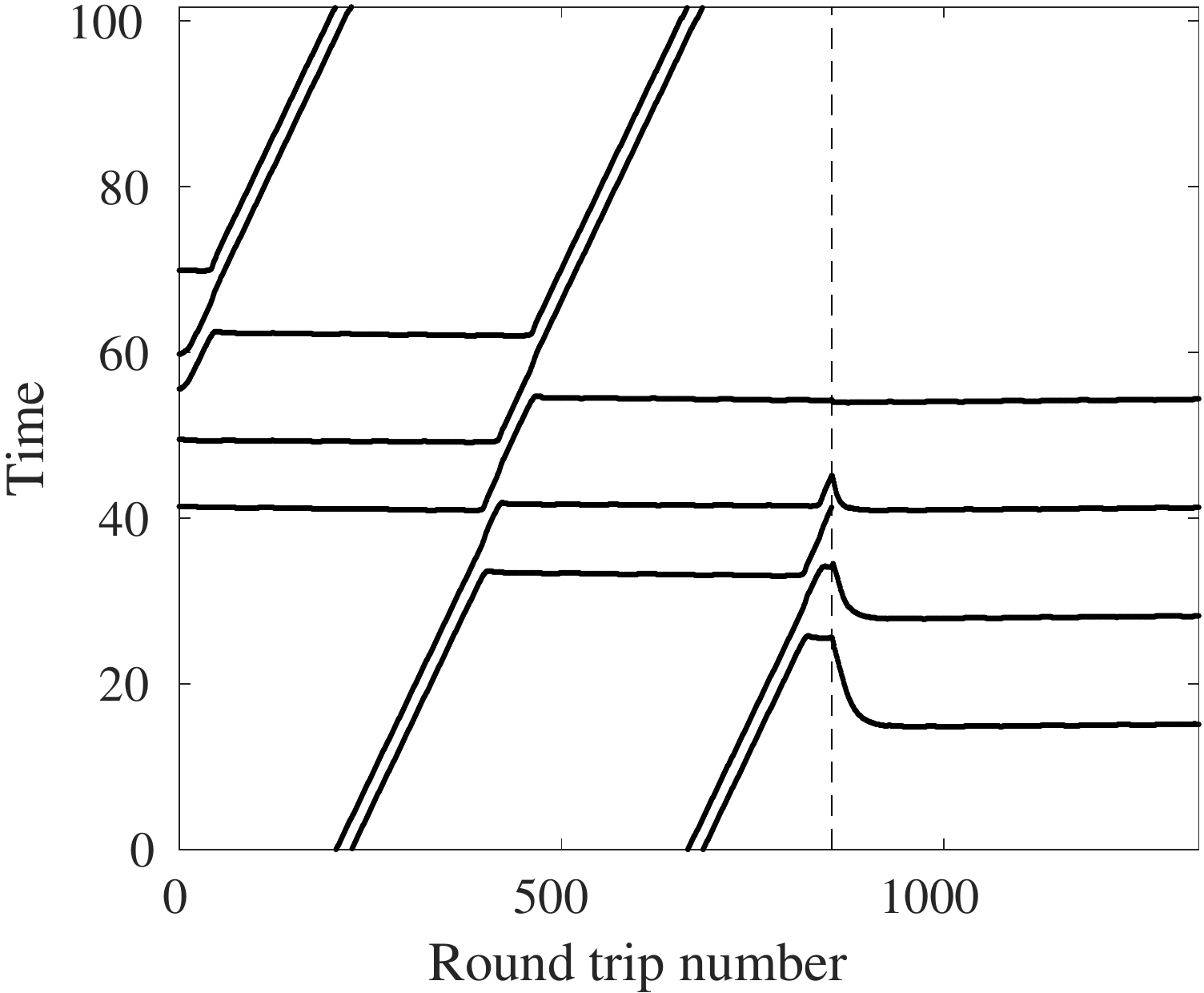}
\caption{Pseudo space-time representation of the positions of multiple LSs
obtained by DNSs of the Adler Eq.~\eqref{eq:adler}. The simulation
starts with five LSs at varying distances. Parameters are $\Delta=1.1$,
$\chi=1.7$, $\chi=0.99$ and $\tau=100$. The parameters change at
the dashed line to $\Delta=0.6$, $\psi=2.1$, $\chi=0.99$ and $\tau=100$.}
\label{fig:multiplelss} 
\end{figure}

In the following we will combine some of the results obtained in the
previous sections to show how one could manipulate a system of multiple
LSs by changing the systems parameters. We start the simulation in
the regime where LSs can form a molecule bound state, if they are
close enough $\left(\Delta,\chi,\psi,\tau\right)=\left(1.1,0.99,1.7,100\right)$
as seen in Fig.~\ref{fig:interaction}(b), with five LSs at varying
starting distances. Two of the LSs are close enough to form a molecule
bound state that starts to drift. Figure \ref{fig:multiplelss} shows
the DNS in pseudo-space-time. Upon reaching another LS the molecule
bound state splits in two while one of the LSs forms another molecule
bound state with the newly encountered LS. Since the domain is periodic
this molecule bound state moves trough it repetitively and interacts
with every other LSs in the delay line. This dynamics is in good agreement
with the experimental results of Fig.~3 of \citep{GJB-CHA-17}. After
round trip $850$ (indicated by a dashed line) the system parameters
were changed to $\left(\Delta,\chi,\psi,\tau\right)=\left(0.6,0.99,2.1,100\right)$
which is close to the parameters in Fig.~\ref{fig:locking} and leads
to the same regular locking behavior. One can clearly see that the
LSs organize themselves in an equidistant pattern since every LS locks
the next one into a minimum of the induced potential. However there
are only four LSs on the right side of the dashed line while the system
started with five LSs prior to the parameter change. The annihilation
of one LS is most likely caused by the abrupt change of parameters,
combined with the small distance with neighbors.

\section{Conclusion}

The time-delayed Adler equation is a prototypical model for the dynamics
of the phase of an injected semiconductor laser with coherent injection
and delayed feedback. It consists in a single, $2\pi$ periodic degree
of freedom and of only three control parameters. It is arguably one
of the simplest model giving rise to topological localized structures.
In this paper we studied the bifurcation mechanisms that govern the
stability of the locked solution and the appearance of stable LSs.
We have found that the locked solution can become unstable via saddle-node
bifurcations, as in the standard, not delayed Adler equation, but
also via Andronov-Hopf bifurcations, which is a direct consequence
of the presence of the time delay. In the long delay limit, approximations
of the SN and AH borders were given using the quasi-continuous spectrum
method. We have found that the branches of single and multiple LSs
are usually connected and that parameter sweeps induce transitions
between states with different numbers of evenly spaced LSs. The branches
of LSs were found to emerge either from diverging period solutions
or from saddle-node of limit cycle bifurcations. Finally, we provided
the derivation of the effective equations of motion governing the
distance between LSs. We have found that the leading eigenvalue expansions
can be obtained solving a linear, time-advanced, equation and that
the lack of parity leads to non-reciprocal interactions and drifting
bound states. Finally, we observed how the transition from real towards
complex eigenvalues explains the creation of stable bound states.
Further works would consider the statistical dynamics of a large ensemble
of LSs.

\section*{Acknowledgments}

J.J. acknowledge the financial support of the MINECO Project MOVELIGHT
(PGC2018-099637-B-100 AEI/FEDER UE). S.G. acknowledges the PRIME program
of the German Academic Exchange Service (DAAD) with funds from the
German Federal Ministry of Education and Research (BMBF).


\end{document}